\documentclass{article}
\usepackage{graphicx}  
\usepackage{amsmath}   
\usepackage[compress]{cite}
\usepackage{amssymb}   
\usepackage{bm} 
\usepackage{dcolumn}
\usepackage{color}
\usepackage{mathrsfs}
\usepackage{amsfonts}
\usepackage{varioref}
\usepackage{textcomp}
\DeclareMathOperator{\sech}{sech}
\RequirePackage[colorlinks,citecolor=blue,urlcolor=magenta,linkcolor=blue]{hyperref}
\allowdisplaybreaks
\usepackage{array}
\renewcommand{\arraystretch}{1.2}
\newcounter{mnotecount}[section]

\renewcommand{\themnotecount}{\thesection.\arabic{mnotecount}}

\newcommand{\mnotex}[1]
{\protect{\stepcounter{mnotecount}}$^{\mbox{\footnotesize
$
\bullet$\themnotecount}}$ \marginpar{
\raggedright\small\em
$\!\!\!\!\!\!\,\bullet$\themnotecount: #1} }

\labelformat{section}{Section #1} 
\labelformat{subsection}{Section #1} 
\labelformat{subsubsection}{Section #1}
\labelformat{subsubsubsection}{Section #1}
\labelformat{equation}{Eq.~(#1)} 
\labelformat{figure}{Fig.~#1} 
\labelformat{subfigure}{Fig.~\thefigure#1} 
\labelformat{table}{Table~#1} 
\labelformat{appendix}{Appendix #1}
\newcommand{\be}{\nopagebreak[3]\begin{equation}}
\newcommand{\ee}{\end{equation}}

\newcommand{\ba}{\nopagebreak[3]\begin{eqnarray}}
\newcommand{\ea}{\end{eqnarray}}

\title{\bf Echoes from braneworld wormholes}
\author{Shauvik Biswas\footnote{intsb6@iacs.res.in}~\footnote{shauvikbiswas2014@gmail.com}$~^{1}$, Mostafizur Rahman\footnote{mostafizur.r@iitgn.ac.in}$~^{2}$ and Sumanta Chakraborty\footnote{sumantac.physics@gmail.com}$~^{1}$
\\
$^{1}${\small{School of Physical Sciences, Indian Association for the Cultivation of Science, Kolkata-700032, India}}\\
$^{2}${\small{Indian Institute of Technology, Gandhinagar, Gujarat-382355, India}}}
\begin{document}
  
\maketitle
\begin{abstract}

We have studied the stability of wormhole geometries, under massless scalar, electromagnetic and axial gravitational perturbations, in the context of higher dimensional spacetimes. Intriguingly, the construction of a wormhole spacetime in the presence of higher dimensions, known as braneworld wormholes, \emph{does not} require the existence of exotic matter fields, unlike the scenario in four spacetime dimensions. Being a nonvacuum spacetime, the effective potential experienced by the axial gravitational perturbation differs considerably from the scenarios involving black holes. In particular, the present work provides one of the first attempts to study the gravitational perturbations of the wormhole spacetimes. Our analysis, involving both analytical and numerical techniques, demonstrates that there are echoes in the time domain signal of all the perturbations and the echo time delay is intimately related to the parameters originating from higher dimensions. Thereby combining the attempt to search for wormholes and extra dimensions, with the existence of gravitational wave echoes. Implications and future directions have also been discussed. 

\end{abstract}
\section{Introduction}

Detections of gravitational waves from the merger of binary black holes \cite{LIGOScientific:2016aoc,LIGOScientific:2017bnn,LIGOScientific:2018mvr,Abbott:2020niy,Abbott:2020jks}, as well as of binary neutron stars \cite{LIGOScientific:2020aai,LIGOScientific:2017vwq} and observation of  the black hole shadow \cite{EventHorizonTelescope:2019ths,EventHorizonTelescope:2019pgp,EventHorizonTelescope:2022xnr} have become the two pillars to understand the nature of gravity in the strong field regimes. Gravitational waves emerging from the collision of binary black holes, depict three phases --- (a) the inspiral phase, this is when the binary black holes approach each other from a large distance and is more or less captured by the post-Newtonian or, the post-Minkowski computations \cite{Blanchet:2013haa,poisson_will_2014,Will:1993ns,Berti:2018cxi}) , (b) the merger phase, when the two black holes merge with each other, leading to an unstable remnant (this can only be worked out using numerical relativity computations, see \cite{PhysRevLett.95.121101,PhysRevD.74.041501,PhysRevLett.111.241104,PhysRevD.93.044006, PhysRevD.100.024021,PhysRevLett.99.181101}) and finally (c) the ringdown phase, where the unstable remnant settles down to a stable black hole configuration, after emitting quasinormal modes (generally, can be understood by the black hole perturbation theory \cite{Teukolsky:1973ha,Chandrasekhar1984,Kokkotas:1999bd,Berti:2009kk,Schutz1985BlackHN,Konoplya:2011qq,Berti:2018vdi}). Instead of a black hole, if we have some exotic compact objects \cite{Cardoso:2019rvt,Maggio:2018ivz,Wang:2019rcf,Mark:2017dnq,Sennett:2017etc,Dey:2020pth,Dey:2020lhq,Abedi:2018npz} (ECOs), whose radii are less than or equal to the photon sphere\cite{Claudel:2000yi}, the inspiral phase will more or less remain the same, except for some change at $(2.5 \times$ log-velocity) PN order, due to modifications in the tidal effects \cite{Henry:2020ski,Narikawa:2021pak,Chakraborty:2021gdf,Chakravarti:2018vlt,Datta:2020rvo}. Due to complications and limited use of numerical relativity, it is yet not possible to simulate the merger state for these exotic compact objects. However, the ringdown phase is of significant interest, since the quasinormal modes (QNMs) depend on the nature of the compact object, through the boundary conditions imposed on the perturbation equations. For black holes, the boundary conditions near the horizon is purely ingoing, while for these compact objects they will have an outgoing part as well. This completely modifies the structure of the QNMs, leading to echoes in the time-domain signal for the black holes. This is one of the most major smoking gun tests for the non-black-hole nature of compact objects \cite{Cardoso:2016rao,Maselli:2017cmm,Chakraborty:2022zlq,Maggio:2020jml,Chakravarti:2021clm,Mukherjee:2022wws,Bhagwat:2019dtm,Pani:2010em,Wang:2018mlp}. 

The major drawback of these models being, they require exotic matter fields \cite{Cardoso:2016rao} and are often unstable under superradiant instability \cite{Maggio:2018ivz}. It seems that the presence of extra dimensions can actually cure these drawbacks, since through the AdS/CFT correspondence \cite{Witten:1998qj,Maldacena:1997re,Emparan:2002px}, it follows that for a four dimensional brane embedded in a five dimensional AdS-like bulk, the brane will inherit quantum corrections \cite{Emparan:1999wa}, which results into a tiny shift in the location of the horizon. This provides a natural model for ECOs \cite{Dey:2020lhq}. Also for rotating braneworld black holes, the presence of the extra dimension significantly reduces the superradiant instability \cite{Dey:2020pth,Biswas:2021gvq}, thereby making these black holes more stable. On the observational side as well, the ringdown phase of the loudest gravitational wave measurement, namely GW150914, is consistent with the braneworld scenario, provided its parameters are within a specified range \cite{Mishra:2021waw}. The observation of black hole shadow, on the other hand, is also fully consistent with the braneworld scenario \cite{Banerjee:2019nnj}. Therefore, the presence of higher dimensions seems consistent with these strong field tests of gravity and possibly even better than the existing models of ECOs with exotic matter. 

\begin{figure}
	\centering
	\minipage{0.48\textwidth}
	\includegraphics[width=\linewidth]{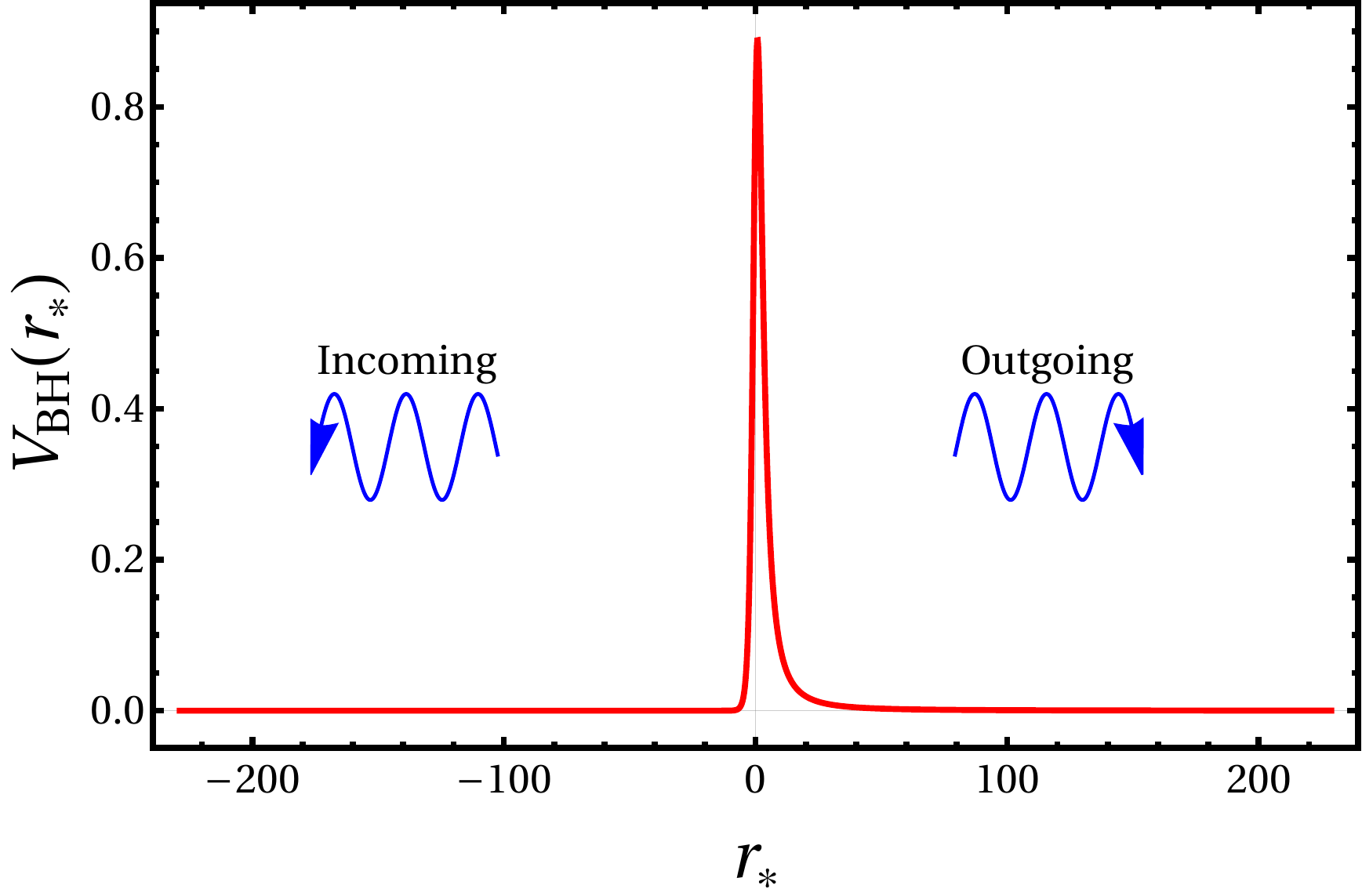}
	\endminipage\hfill
	\minipage{0.48\textwidth}
	\includegraphics[width=\linewidth]{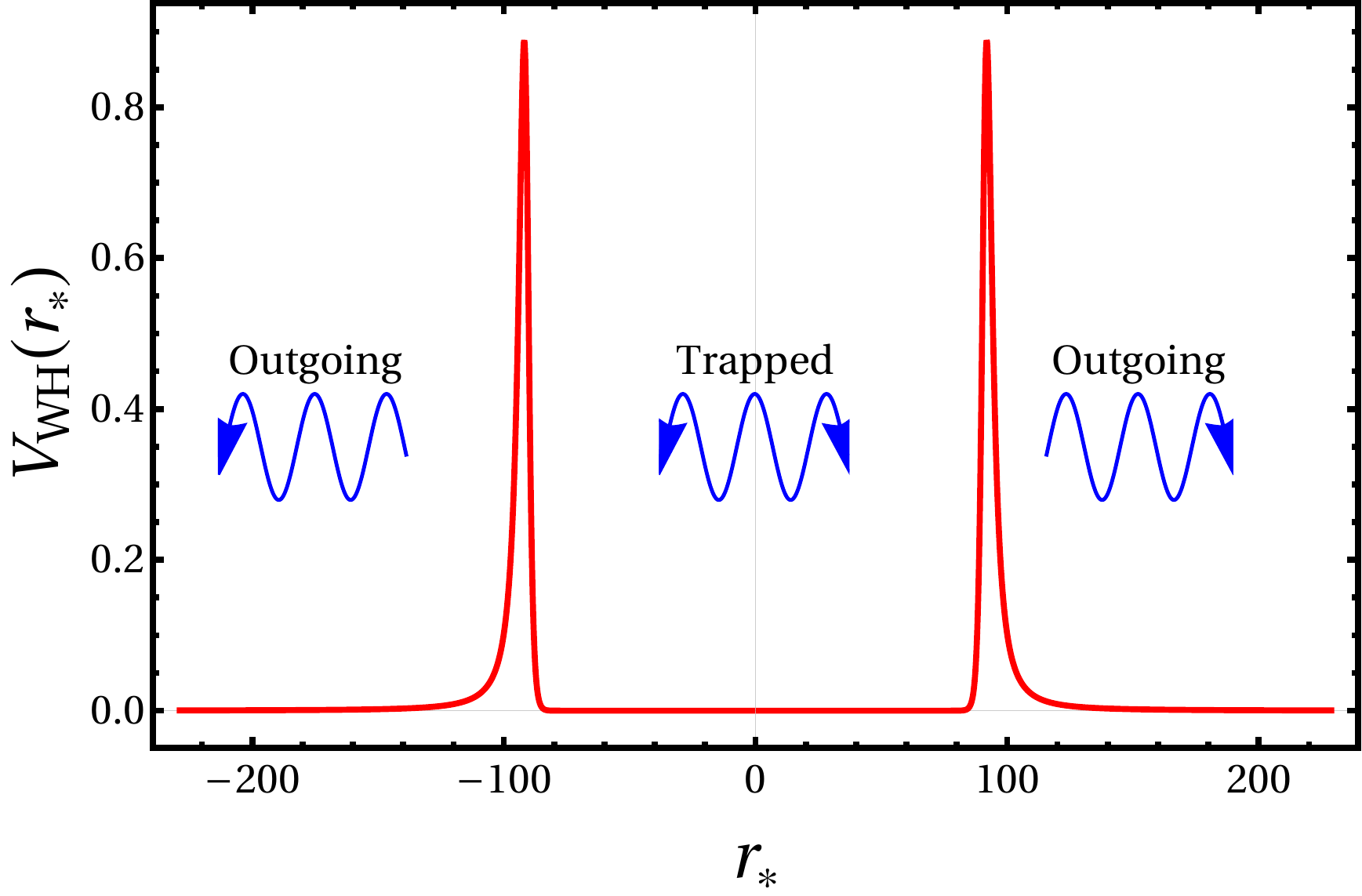}
	\endminipage
	\caption{A typical representation of the effective perturbation potential as a function of the tortoise coordinate $r_*$ is presented for a black hole (left panel) and a wormhole (right panel). For the black hole case, the single-bump potential approximately peaks around the photon sphere of the black hole. However, the wormhole potential is a mirror-symmetric (about $r_*=0$) double-bump potential. The black hole QNMs satisfy the incoming boundary condition at the horizon and outgoing boundary condition at the infinity, whereas the wormhole QNMs satisfy the outgoing boundary condition at both the infinities. Part of the wave is quasi-trapped inside the double-bump potential, back and forth motion of which gives rise to the echo signal.}\label{fig_WH_BH}
\end{figure}

Among ECOs, the most useful and intriguing ones are the wormholes\cite{Morris:1988cz,Morris:1988tu,Damour:2007ap,DuttaRoy:2019hij,2017,Lobo:2020ffi,visser1995lorentzian}. Since wormholes connect two distinct universes through a throat, the fact that the wormhole spacetime will have a reflectivity is obvious. This is because, waves from our universe can go into the other universe and get reflected by the photon sphere\cite{Bueno:2017hyj,Yang:2021cvh} of the other universe and reemerges in our universe as if it has been reflected by the wormhole throat, leading to echoes (see \ref{fig_WH_BH}). However, in order to have traversable wormholes --- central to the emergence of echoes from wormholes --- one needs exotic matter \cite{Morris:1988cz}. Again, extra dimensions come to the rescue, since extra dimensions can sustain braneworld wormholes \emph{without} exotic matter \cite{Kar:2015lma,PhysRevD.65.084040,Bronnikov:2002rn,Bronnikov:2003gx}, while the contribution of the extra dimension itself on the brane will appear exotic. This appears to be an interesting proposal, where we can construct a wormhole with normal matters and such static and spherically symmetric wormhole solution already exists in the literature \cite{Dadhich:2001fu}. In this work, we wish to study the stability of the wormhole solution under scalar, electromagnetic and gravitational perturbations (for an earlier attempt in this direction, see \cite{Aneesh:2018hlp}). In particular, we wish to explore the late time echoes present in the time domain waveform of the perturbed braneworld wormhole.    

The paper is organized as follows: We start with a brief review of the wormhole solution in the braneworld context in \ref{wormhole_review} and subsequently we present various properties of the wormhole solution in \ref{wormhole_properties}. The perturbation of the braneworld wormhole under scalar, electromagnetic and gravitational perturbation and the resulting master equations have been presented in \ref{wormhole_perturb}, with the numerical solutions depicting the quasinormal modes and the time-domain waveform in \ref{wormhole_qnm}. Finally, we comment on the possibility of obtaining a rotating braneworld wormhole, starting from the static and spherically symmetric one, in \ref{wormhole_rotation} and then we conclude in \ref{wormhole_discussion} with a discussion of the results and on future prospects. 

\emph{Notations and Conventions:} Throughout this paper, we will use the positive signature convention, such that the flat metric in four spacetime dimensions can be expressed as, $\eta_{\mu \nu}=\textrm{diag}(-1,+1,+1,+1)$. We will use the Greek indices $\mu,\nu,\rho,\cdots$, in order to describe four-dimensional spacetime coordinates. We will also set the fundamental constants, $G$ and $c$ to be unity.

\section{A brief review of the low energy effective action on the brane and the associated wormhole solution}\label{wormhole_review}

As mentioned in the Introduction, we wish to study the perturbation of the wormhole solution on the brane and hence establish the possible presence of echoes in the ringdown waveform. Before going into the details of the perturbation equations and the numerical techniques thereof, we would first like to briefly review the origin of the wormhole solution itself in order to set the perspective. We will closely follow the analysis of \cite{Kanno:2002ia} in order to expand the bulk (five dimensional) geometrical entities as a power series in the ratio of (bulk/brane) curvature length scales. This results into gravitational field equations on the brane containing only local quantities, unlike \cite{PhysRevD.62.024012}. 

In this scheme of obtaining local gravitational field equations on the brane, one starts by considering an $\textrm{AdS}_{5}$ bulk spacetime with a compact spacelike extra dimension (the extra coordinate is being denoted as $y$), with the two 3-branes located at $y=0$ (denoted as A, Planck brane) and $y=\ell$ (denoted as B, visible brane). The bulk spacetime is described by the following metric ansatz,
\begin{align}\label{metric-ansatz}
ds^{2}=e^{2\phi(x)}dy^{2}+\widetilde{g}_{\mu\nu}(y,x)dx^{\mu}dx^{\nu}~,
\end{align}
where, $x$ denotes the brane coordinates collectively and $\phi(x)$ is called the radion field, describing the inter-brane separation $d(x)$, given by, $d(x)\equiv e^{\phi(x)}\ell$. Note that this interbrane separation, being a function of the brane coordinates, behaves as an extra field for the brane observers with interesting phenomenology \cite{Chakraborty:2015zxc,Chakraborty:2013ipa,Chakravarti:2019aup}. 

Afterwards one expands the bulk Einstein's equations as a power series in $(\ell/\textrm{brane~curvature~scale})$, resulting in two sets of Einstein's equations, one on the Planck brane and another on the visible brane. These two equations depend on the first order term of the electric part of the projected bulk Weyl tensor, whose elimination yields the following effective gravitational field equations on the visible brane,
\begin{align}\label{eom-B}
G_{\mu\nu}=\frac{\kappa^{2}}{\ell \Phi}T^{\rm B}_{\mu\nu}+\frac{\kappa^{2}(1+\Phi)}{\ell \Phi}T^{\rm A}_{\mu\nu}+\frac{1}{\Phi}T^{\Phi}_{\mu\nu}~,
\end{align}
where, 
\begin{align}\label{energy-Phi}
T^{\Phi}_{\mu\nu}=\bigg(\nabla_{\mu}\nabla_{\nu}\Phi-g_{\mu\nu}\nabla^{\alpha}\nabla_{\alpha}\Phi\bigg)-\frac{3}{2(1+\Phi)}\bigg(\nabla_{\mu}\Phi\nabla_{\nu}\Phi-\frac{1}{2}g_{\mu\nu}\nabla_{\alpha}\Phi\nabla^{\alpha}\Phi\bigg)\quad\text{with}\quad \Phi=\exp[2e^{\phi(x)}]-1~,
\end{align}
along with $T^{\rm A}_{\mu \nu}$ being the energy-momentum tensor on the Planck brane and $T^{\rm B}_{\mu \nu}$ is the energy-momentum tensor on the visible brane. Note that here the covariant derivative $\nabla_{\alpha}$ is with respect to the visible brane metric $g_{\mu\nu}$, $\kappa^{2}$ is the five dimensional gravitational constant and hence the ratio $(\kappa^{2}/\ell)$ must play the role of four dimensional gravitational constant, and, finally, in the above equation $\Phi$ is the nonlinear realization of the radion field $\phi$ on the visible brane. Similarly, we obtain the field equation for $\Phi(x)$, the four dimensional incarnation of the radion field as
\begin{align}\label{EOM-Phi}
\nabla^{\alpha}\nabla_{\alpha}\Phi=\frac{\kappa^{2}}{\ell}\frac{T^{\rm A}+T^{\rm B}}{2\omega+3}-\frac{1}{2\omega+3}\frac{d\omega}{d\Phi}(\nabla^{\alpha}\Phi)(\nabla_{\alpha}\Phi)~\quad\text{with}\quad\omega=-\frac{3\Phi}{2(1+\Phi)}~,
\end{align}
where $T^{\rm A}$ and $T^{\rm B}$ are the traces of the energy momentum tensor on the Planck and on the visible brane, respectively. Despite the striking similarity of the above equation with the one in the Brans-Dicke theory \cite{PhysRev.124.925}, in the present scenario we have extra contributions from the energy-momentum tensors of the branes, as well as, in this case, $\omega$ becomes a function of the radion field incarnation $\Phi$. The gravitational field equations on the visible brane in \ref{eom-B} together with \ref{energy-Phi} and \ref{EOM-Phi} complete the picture of a low energy effective scalar-tensor theory on the brane, provided the energy-momentum tensors on Planck and visible branes are known. 

The most simplest choice, under which a nontrivial solution of the gravitational field equations in \ref{eom-B} can be obtained, corresponds to that of vacuum Planck brane, with $T^{\rm A}_{\mu \nu}=0$. While for the visible brane one considers, $T_{\mu \nu}^{\rm B}$ to be made out of anisotropic perfect fluid, such that,
\begin{align}\label{mattar-visible}
T^{\mu\nu}_{\rm B}=(\rho+p)u^{\mu}u^{\nu}+pg^{\mu\nu}-(p-\tau)w^{\mu}w^{\nu}~,
\end{align}
where $u^{\mu}$ is the timelike fluid four-velocity and $w^{\mu}$ represents the spacelike vector transverse to $u^{\mu}$, satisfying $w_{\nu}w^{\nu}=1$ and $w^{\nu}u_{\nu}=0$. In the above $\rho$ represents the energy density, $p$ represents the pressure transverse to $w^{\mu}$ and $\tau$ represents the pressure along $w^{\mu}$. Given this energy momentum tensor on the brane, one solves the Einstein's equations for the static and spherically symmetric metric from \ref{eom-B}. For this purpose, note that the trace of $T^{\Phi}_{\mu \nu}$ identically vanishes \cite{Shaikh:2016dpl} and hence the Ricci scalar of the visible brane gets related to the trace of the energy-momentum tensor on the visible brane. Imposing the condition that the Ricci scalar of the visible brane should vanish further simplifies the scenario and results in the following condition, $-\rho+\tau+2p=0$, which also has interesting astrophysical consequences, see \cite{bondi1992anisotropic}. As a consequence of the vanishing Ricci scalar, in addition to the Einstein's equations, we get an additional relation between the $g_{tt}$ and the $g^{rr}$ components of the static and spherically symmetric metric. Further, owing to the traceless-ness condition satisfied by the brane matter, the field $\Phi$ is independent of the energy momentum components of the brane matter, as evident from \ref{EOM-Phi}. Thus we have six unknown variables, the matter energy momentum tensor components $(\rho,p,\tau)$, the metric components, $g_{tt}$, $g^{rr}$, and the radion field $\Phi$. However, we have five equations to solve them-(a) two Einstein's equations, (b) one conservation equation for the brane energy momentum tensor,  (c) the relation for vanishing Ricci scalar, and (d) the field equation for the radion field. Since we need one supplementary condition, following \cite{Dadhich:2001fu}, we assume the following line element,
\begin{align}\label{S-kar-form}
ds^{2}=-\bigg(\alpha+\lambda\sqrt{1-\frac{2M}{r}}\bigg)^{2}dt^{2}+\frac{dr^{2}}{1-\frac{2M}{r}}+r^{2}(d\theta^{2}+\sin^{2}\theta d\phi^{2})~,
\end{align}
which satisfies the vanishing Ricci scalar equation. Here $\alpha$ and $\lambda$ are two constants, taken to be real and positive \cite{Kar:2015lma}, in order to avoid any formation of a naked singularity. Note that for $\alpha=0$ and $\lambda=1$, this solution reduces to the standard Schwarzschild solution. The field $\Phi$ can also be solved from \ref{EOM-Phi}, as a function of the radial coordinate $r$, however, the solution is easier to write down in the isotropic coordinate $r^{\prime}$, which is related to Schwarzschild coordinate $r$ by $r\equiv r^{\prime}[1+(M/2r^{\prime})]^{2}$, such that, 
\begin{align}
\sqrt{1+\Phi(r)}=\frac{C_{1}}{M\lambda}\ln \left[\frac{2qr'+M}{2r'+M}\right]+C_{4}~,
\label{Sol-Phi}
\end{align}
where, $q\equiv [(\alpha+\lambda)/(\alpha-\lambda)]$ and $C_{1}$, $C_{4}$ are the constants of integration. Note that for $\alpha<\lambda$, one defines $q=[(\alpha+\lambda)/(\lambda-\alpha)]$, such that in the Schwarzschild limit, $q=1$. Therefore, the field $\Phi$ is a constant in the Schwarzschild background, which can as well be taken to be zero. Finally, given the line element in \ref{S-kar-form} and the radion field $\Phi$ in \ref{Sol-Phi}, the components of the brane energy momentum tensor can be obtained from the Einstein's equations, whose expressions can be found in \cite{Kar:2015lma}. 
 
Further, to sustain various wormhole solutions in general relativity, a violation of the energy condition is bound to appear. In particular, near the throat, we cannot avoid the violation of the weak energy condition\cite{visser1995lorentzian}, such that to some observers, the energy density will appear to be negative. In contrast to the above, for the braneworld case, the effective energy momentum tensor is a combination of the brane energy momentum tensor, along with an energy-momentum tensor for the radion field. Therefore, even if the combined effective energy momentum tensor of the brane matter and the radion field violate energy conditions, it is always possible to avoid the violation of energy conditions in the matter sector, such that all such violations appear in the radion sector alone. This is how in the braneworld scenario it is possible to have a wormhole \emph{without} any exotic matter field on the brane.  Having described the details of the wormhole solution on the brane, in what follows we will discuss some other forms of the above metric and its properties, which will be useful for our purposes. 

\section{The wormhole solution and its various characteristics}\label{wormhole_properties}

Having outlined the derivation of the solution corresponding to the braneworld wormhole explicitly, let us explore its various properties. In particular, we are interested in the parameter region, where $\alpha\gtrsim 0$ and $\lambda\simeq 1$, i.e., a continuous limit to the Schwarzschild geometry can be taken. From \ref{S-kar-form} it is easy to see that any $r=$~constant hypersurface becomes null at $r=2M$. This is because $g^{rr}$ identically vanishes when $r=2M$. However, unlike the Schwarzschild solution, $r=2M$ does not represent an event horizon in the present scenario. To justify this, we can see that the timelike Killing vector field $\xi_{t}^{\mu}=(\partial/\partial t)^{\mu}$ does not coincide with the null vector at $r=2M$. Moreover, $\xi_{t}^{2}=-\alpha^{2}$ at $r=2M$. Therefore $\xi_{t}^{\mu}$ remains timelike even at $r=2M$. This explicitly demonstrates that $r=2M$ is not a causal barrier for the observers who reside outside $r=2M$. This confirms that, for $\alpha\neq 0$ the spacetime described by the line element in \ref{S-kar-form} does not exhibit any event horizon\cite{vishveshwara1968generalization}. In particular, it depicts a wormhole with its throat at $r=2M$. This can also be seen using the embedding diagram \cite{Morris:1988cz}. Further, the metric is regular at $r=2M$, since the evaluation of the Kretschmann scalar $R_{\mu \nu \alpha \beta}R^{\mu \nu \alpha \beta}$ yields
\begin{align}\label{K-Scalar}
R_{\mu\nu\rho\sigma}R^{\mu\nu\rho\sigma}=\frac{24 M^{2}}{r^{6}}\left[1+\left(1-\frac{2M}{r}\right)\left(\frac{\alpha}{\lambda}+\sqrt{1-\frac{2M}{r}}\right)^{-2}\right]~.
\end{align}
As evident from the above expression, in the limit $r\rightarrow 2M$, the Kretschmann scalar becomes $(3/8M^{4})$, exactly \emph{half} of the corresponding value for Schwarzschild. Also, there is no coordinate singularity at $r=2M$, since on any $r=\textrm{constant}$ surface the $dr^{2}$ term does not contribute and all the other metric elements are finite at $r=2M$, as evident from \ref{S-kar-form}. To see more explicitly the properties associated with the surface $r=2M$, we consider a coordinate transformation of the form, $r=2M+x^{2}$ and compute the surface stress tensor describing the jump at the surface $x=0$, caused by the square root term in the $g_{tt}$ component of the metric in \ref{S-kar-form}. As evident, the surface stress tensor satisfies the null energy condition and does not inhibit any exotic matter field, see \ref{appB} for an explicit computation.

At this outset, we would also like to emphasize that the metric as written in \ref{S-kar-form} does not reduce to $\eta_{\mu \nu}$ in the limit $r\rightarrow \infty$, though all the Riemann tensor components vanish in this limit. In order to transform the metric to a manifestly asymptotic flat form (i.e., as $r\rightarrow \infty$, we get $g_{\mu \nu}\rightarrow \eta_{\mu \nu}$), we redefine the time coordinate as, $t\rightarrow (\alpha+\lambda)^{-1}t$, then we obtain
\begin{align}\label{Sumanta-Form}
ds^{2}=-(p+1)^{-2}\bigg(p+\sqrt{1-\frac{2M}{r}}\bigg)^{2}dt^{2}+\left(1-\frac{2M}{r}\right)^{-1}dr^{2}+r^{2}(d\theta^{2}+\sin^{2}\theta d\phi^{2})~\quad\text{with}\quad p\equiv\frac{\alpha}{\lambda}~,
\end{align}
which indeed reduces to $\eta_{\mu\nu}$ at spatial infinity. From our discussion above, it is clear that for the parameter regime of our interest, namely, $M>0$ and $p\gtrsim 0$, the spacetime remains regular, including the throat of the wormhole, located at $r=2M$.

Having described some of the basic properties of the wormhole metric in the context of a two-brane system, let us briefly point out the existence of similar wormhole solutions in the context of a single brane scenario. In the single brane model, instead of the radion field $\Phi$, the gravitational field equations involve the projected bulk Weyl tensor, which is also traceless. Thus in the single brane model as well one can derive wormhole solutions with $R=0$ \cite{PhysRevD.65.084040,Bronnikov:2002rn}, without exotic matter on the brane, however, the interpretation is different and so are the solutions. For example, in the present scenario, there is a clear interpretation of the departure of the wormhole solution from the Schwarzschild spacetime- nontrivial dependence of the interbrane separation on the radial coordinate. While such a clear interpretation cannot be arrived at from the solutions in \cite{PhysRevD.65.084040,Bronnikov:2002rn}. On the other hand, the results presented here hold true only in the perturbative regime of the bulk to brane curvature scales, but the single brane scenario, as presented in \cite{PhysRevD.65.084040,Bronnikov:2002rn} holds in the nonperturbative regimes as well. In what follows, we will concentrate on the wormhole solution in the two brane system and shall consider the single brane extension elsewhere.

The above redefinition of the time coordinate have reduced the braneworld wormhole solution into a two parameter ($p$,$M$) family of wormhole spacetimes. For $p=0$, it reduces to the standard Schwarzschild solution, while for $p\neq 0$, this depicts a wormhole solution, whose stability will be studied in this work. From now on we will use the metric in \ref{Sumanta-Form} for our subsequent analysis. For the future, it is useful to define a tortoise coordinate $r_{*}$ for the above wormhole metric as
\begin{align}\label{tortoise-1}
dr_{*}=\frac{dr}{y(r)}~\quad\text{with}\quad y(r)\equiv \frac{{\bigg(p+\sqrt{1-\frac{2M}{r}}\bigg)\sqrt{1-\frac{2M}{r}}}}{(p+1)}~,
\end{align}
which upon integration yields\cite{Aneesh:2018hlp}
\begin{align}\label{tortoise-2}
r_{*}=(p+1)M\left[\frac{2(p-\beta)(2p-\beta)}{(p^{2}-1)[(p-\beta)^{2}-1]}+4\frac{\ln\frac{\beta}{p}}{(p^{2}-1)^{2}}+\frac{(p-2)\ln(1-p+\beta)}{(p-1)^{2}}-\frac{(2+p)\ln(1+p-\beta)}{(1+p)^{2}}\right]~,
\end{align}
where $\beta\equiv p+\sqrt{1-(2M/r)}$. In arriving at \ref{tortoise-2} we have chosen the integration constant such that the tortoise coordinate $r_{*}$ vanishes at $r=2M$ (equivalently, at $\beta=p$), such that in terms of the tortoise coordinate, the wormhole throat is located at $r_{*}=0$. Since any wormhole spacetime can be viewed as if two black hole spacetimes have been joined together at the throat \cite{Mark:2017dnq,Cardoso:2016rao}, therefore by this choice of the tortoise coordinate $r_{*}$, we can describe the entire spacetime by extending the range of the tortoise coordinate to $r_{*}\in(-\infty,+\infty)$. This makes $(t,r_{*},\theta,\phi)$ as the global coordinate patch for the above wormhole spacetime. Moreover, the tortoise coordinate helps to reduce the linear perturbation equations to that of the one-dimensional Schr\"{o}dinger equation, thereby leading to simpler numerical routines to solve for the quasinormal modes, as we will demonstrate in the subsequent sections.  

At this outset, we wish to briefly talk about one of the most important parameters in our model, namely the throat length of the braneworld wormhole. This is best understood by expressing the tortoise coordinate defined in \ref{tortoise-2} for the braneworld wormhole in terms of the tortoise coordinate for the Schwarzschild black hole, and corrections to it due to nonzero values of $p$ ($\gtrsim 0$), yielding (for a derivation, see \ref{AppA})
\begin{align}\label{tortoise-3}
r_{*}\simeq r^{\rm BH}_{*}+\frac{L}{2}+p\bigg[(r-2M)+2M\ln{\left(\frac{r}{2M}-1\right)}-&r\sqrt{1-\frac{2M}{r}}+\frac{4M}{\sqrt{1-\frac{2M}{r}}}\nonumber\\
&-3M \ln{\left(\frac{1+\sqrt{1-\frac{2M}{r}}}{1-\sqrt{1-\frac{2M}{r}}}\right)}\bigg]-4Mp\ln{ p}~.
\end{align}
Here, $r^{\rm BH}_{*}$ corresponds to the tortoise coordinate of the Schwarzschild black hole, which reads, $r^{\rm BH}_{*}\equiv r+2M\ln|(r/2M)-1|$, the length $L\equiv -4M(1+2\ln p)$ (which is a positive quantity, as $p<1$) and then we have terms $\mathcal{O}(p)$ and of $\mathcal{O}(p\ln p)$, respectively. We would like to emphasize that the tortoise coordinate $r_{*}$ is regular everywhere, since the radial coordinate $r$ is also regular. In particular, one can check that $r_{*}(2M)=0$, unlike the case of Schwarzschild black hole. This is because, the $r=2M$ is not a coordinate singularity. If we now neglect all the terms proportional to $p$ in \ref{tortoise-3}, we will get
\begin{align}\label{tortoise-4}
r_{*}=r^{\rm BH}_{*}+\frac{L}{2}~. 
\end{align}
It is generally argued \cite{Bueno:2017hyj} that $L$ should represent the distance between the maxima of the two photon spheres located on both sides of the throat and is referred to as the throat length. To see this explicitly one may take a different route, first of all, the photon sphere in the present scenario must be located at $r_{\rm ph}=3M+\mathcal{O}(p)$ and hence the distance between the photon spheres, on both sides of the throat, should have the following expression,
\begin{align}
\bar{L}=2r_{*}(3M)=-4M\left[\ln\sqrt{\frac{4}{e}}+2\ln p\right]~,
\end{align}
where, $r_{*}$ is given by \ref{tortoise-2} and terms of $\mathcal{O}(p)$ have been neglected. Note that $L$ and $\bar{L}$ are both identical except for some small differences, and hence we can as well consider $L$ as the throat length.

The existence of a nonzero throat length is crucial for the existence of echoes in the late time signal from a perturbed braneworld wormhole. This is because a part of the primary signal generated near one of the photon spheres, due to external perturbation, will traverse through the throat and will be reflected by the photon sphere potential on the other side of the throat, which will reappear in our universe through the wormhole throat. Thus to the asymptotic observer, these reflected signals will appear as echoes of the primary signal. This discussion shows that nonvanishing throat length $L$ plays the key role in producing the echoes of the primary signal. Therefore, the existence of a nonzero throat length is essential for the braneworld wormhole to behave like a black hole mimicker.

\section{Linear perturbation of the wormhole solution}\label{wormhole_perturb}

In this section, we will present the master equations satisfied by linear perturbations, due to scalar, electromagnetic and gravitational fields in the background of the braneworld wormhole solution considered above. In particular, we will demonstrate how all of these perturbations differ from that of the Schwarzschild black hole, for a nonzero value of the parameter $p$. We start with the simplest case of a massless scalar perturbation and then shall discuss the electromagnetic perturbation, before finalizing with the gravitational perturbations of the wormhole geometry. 

\subsection{Massless scalar perturbation}

Let us consider how massless scalar perturbations evolve in the background geometry of the braneworld wormhole, expressed in \ref{Sumanta-Form}. We will content ourselves with linear perturbations alone, that is the perturbing scalar field (let us denote it as $\Psi(x)$) is much smaller of the background metric components and the radion field (or its nonlinear realization $\Phi$). In this scenario, the perturbing field $\Psi(x)$ satisfies the Klein-Gordon equation in the wormhole background,
\begin{align}\label{EOM-psi}
g^{\mu \nu}\nabla_{\mu}\nabla_{\nu}\Psi(x)=0~,
\end{align}
where the covariant derivative $\nabla_{\mu}$ is with respect to wormhole metric $g_{\mu \nu}$, described in \ref{Sumanta-Form}. Since the background spacetime is spherically symmetric and static we can decompose the perturbing scalar field as,
\begin{align}\label{psi-ansatz}
\Psi(t,r,\theta,\phi)=\sum_{l=0}^{\infty}\sum_{m=-l}^{l}\frac{\psi_{lm}^{(0)}(r)}{r}e^{-i\omega t}Y_{lm}(\theta,\phi)~.
\end{align}
Substitution of the above ansatz for the perturbing scalar field $\Psi$ in \ref{EOM-psi}, describing its evolution, yields the following equation for the radial part $\psi_{lm}^{(0)}(r)$ of the perturbation, 
\begin{align}\label{radial-master}
\frac{d^{2}\psi_{lm}^{(0)}}{dr_{*}^{2}}+\Big[\omega^{2}-V_{l}^{(0)}(r)\Big]\psi^{(0)}_{lm}=0~,
\end{align} 
where $r_{*}$ corresponds to the tortoise coordinate defined in \ref{tortoise-1}. For a generic static and spherically symmetric background metric of the form $\textrm{diag.}[-f(r),\{1/h(r)\},r^{2},r^{2}\sin^{2}\theta]$ (with $f(r)\neq h(r)$), the potential $V_{l}^{(0)}(r)$, appearing in the radial perturbation equation of a massless scalar field becomes
\begin{align}\label{sclar-pot}
V_{l}^{(0)}(r)=f(r)\frac{l(l+1)}{r^{2}}+\frac{\sqrt{fh}}{r}\partial_{r}\left(\sqrt{fh}\right)~.
\end{align}
Comparing, $f(r)$ with the $g_{tt}$ component and $h(r)$ as the $g^{rr}$ component of the braneworld wormhole metric, presented in \ref{Sumanta-Form}, we obtain the following potential, 
\begin{align}\label{Eff-Poten}
V_{l}^{(0)}(r)=\frac{1}{(p+1)^{2}}\Bigg[\frac{M}{r^{3}}\sqrt{1-\frac{2M}{r}}\left(p+\sqrt{1-\frac{2M}{r}}\right)+\frac{1}{r^{2}}\left(p+\sqrt{1-\frac{2M}{r}}\right)^{2}\left(\frac{M}{r}+l(l+1)\right)\Bigg]~,
\end{align}
governing the master equation for the massless scalar perturbation of the background wormhole spacetime. 

As evident, using the ansatz of \ref{psi-ansatz}, motivated by the static and spherically symmetric background spacetime, the Klein-Gordon equation of the massless scalar perturbation reduces into a one dimensional time independent Schr\"odinger-like equation with real potential. We wish to solve this differential equation using appropriate numerical techniques, akin to those in quantum mechanics, which we will elaborate in the next section. 

\subsection{Electromagnetic perturbation}

Alike the case of massless scalar perturbation, the evolution of electromagnetic perturbation can be understood by solving Maxwell's equations in the background of braneworld wormhole. Explicitly, the evolution of the electromagnetic perturbation $A_{\mu}$, is governed by
\begin{align}\label{Maxwell-equation}
\nabla_{\nu}F^{\mu\nu}=0~,
\end{align}
where, $F_{\mu\nu}=\partial_{\mu}A_{\nu}-\partial_{\nu}A_{\mu}$ represents Maxwell's field tensor and the covariant derivative $\nabla_{\mu}$ is with respect to the background braneworld wormhole metric. Given the static and spherically symmetric background spacetime, we can decompose the vector potential $A_{\mu}$ using four dimensional vector spherical harmonics \cite{zerilli1970tensor}, such that,
\begin{align}\label{A-decomp}
A_{\mu}(t,r,\theta,\phi)=\int d\omega \sum_{l,m}\Bigg[a_{lm}(r)e^{-i\omega t}\begin{pmatrix}
0\\
0\\
\frac{1}{\sin\theta}\partial_{\phi}Y_{lm}(\theta,\phi)\\
-\sin\theta\partial_{\theta}Y_{lm}
\end{pmatrix}_{\rm odd}+
e^{-i\omega t}\begin{pmatrix}
f_{lm}(r)Y_{lm}\\
u_{lm}(r)Y_{lm}\\
k_{lm}(r)\partial_{\theta}Y_{lm}\\
k_{lm}(r)\partial_{\phi}Y_{lm}
\end{pmatrix}_{\rm even}\Bigg]~.
\end{align}
In the above, the first term on the right-hand side with the radial dependence as $a_{lm}(r)$ has parity $(-1)^{l+1}$ and hence is referred to as the \emph{odd parity}/\emph{axial} term, while the second term has parity $(-1)^{l}$ and hence is called as the \emph{even parity}/\emph{polar} term. As usual, $l$ represents the angular momentum index and $m$ represents the azimuthal index. Note that there are four unknown radial functions in the above decomposition, $a_{lm}(r)$, $f_{lm}(r)$, $u_{lm}(r)$ and $k_{lm}(r)$, respectively. However, both the axial and polar decomposition of the vector potential, as in \ref{A-decomp}, provides an identical master equation, which reads,
\begin{align}
\frac{d^{2}\psi_{lm}^{(1)}}{dr^{2}_{*}}+\left[\omega^{2}-\underbrace{\frac{l(l+1)}{r^{2}}\frac{\bigg(p+\sqrt{1-\frac{2M}{r}}\bigg)^{2}}{(p+1)^{2}}}_{\equiv V_{l}^{(1)}(r)}\right]\psi^{(1)}_{lm}=0~,
\end{align}
where, the master radial function $\psi^{(1)}_{lm}(r)$, can be expressed in terms of the four unknown radial functions described above, as,\footnote{Note that we have written the master variable for any static spherically symmetric background geometry of the form  $g^{0}_{\mu\nu}=\textrm{diag.}(-f(r),(1/g(r)),r^{2},r^{2}\sin^{2}\theta)$.}
\begin{align}
\psi^{(1)}_{lm}=
\begin{cases}
\frac{r^{2}}{l(l+1)}\sqrt{\frac{g(r)}{f(r)}}  \left(-i\omega u_{lm}-\frac{df_{lm}}{dr}\right)&\text{for}~\text{even parity}
\\
a_{lm}&\text{for}~\text{odd parity}~.
\end{cases}
\end{align}
Note that the $k_{lm}(r)$ term does not appear in the final dynamical equation and there are only two degrees of freedom, one for axial and the other for polar perturbations. This is consistent with the result that a photon has only two independent polarization states. Intriguingly, alike the case of a massless scalar field, for the electromagnetic field as well, the final master equation for radial perturbation resembles the Schr\"{o}dinger equation, but with a different potential. We will see that the same trend will continue to hold true for gravitational perturbations as well. 

\subsection{Axial gravitational perturbation}

Just as the case of electromagnetic perturbation can be decomposed into axial and polar parts, the gravitational perturbation can also be decomposed into these two branches. Since, in general, there are ten independent components of the gravitational perturbation, they separate into three axial and seven polar perturbations. Explicitly, 
\begin{align}
g_{\mu\nu}(t,r,\theta,\phi)&=g^{0}_{\mu\nu}(t,r,\theta,\phi)+h_{\mu\nu}(t,r,\theta,\phi)
\nonumber
\\
&=g^{0}_{\mu\nu}(t,r,\theta,\phi)+h^{\rm axial}_{\mu\nu}(t,r,\theta,\phi)+h^{\rm polar}_{\mu\nu}(t,r,\theta,\phi)~,
\end{align}
where $g^{0}_{\mu\nu}$ represents the background braneworld wormhole spacetime and $h_{\mu\nu}$ represents the gravitational perturbation, which has been decomposed into axial and polar parts. Further, using the gauge degrees of freedom associated with the diffeomorphism invariance \cite{PhysRev.108.1063}, we can reduce the number of axial degrees of freedom to two and the number of polar degrees of freedom to four. Therefore, the polar gravitational perturbation not only has a larger number of degrees of freedom, but the associated perturbation equations are also complex and it is not very clear if one can reduce the system of equations into a single master equation. Given these complications, we will restrict ourselves to the case of axial gravitational perturbation alone. 

Having described the nature of the perturbation, let us move forward and determine the dynamical equation it satisfies. This must be derivable from the computation of $\delta G_{\mu \nu}$, while keeping terms up to linear order in $h_{\mu\nu}$. However, unlike the case of vacuum Einstein's equations, where the perturbations would simply satisfy $\delta G_{\mu \nu}=0$, in the present scenario, there will be nontrivial source terms. For example, both the radion energy momentum tensor $T^{\Phi}_{\mu \nu}$ and the brane energy momentum tensor $T^{\rm B}_{\mu \nu}$ depend on the brane metric $g_{\mu \nu}$ and hence will be perturbed as the brane metric is also perturbed. In addition, there will be perturbation in the radionlike field $\Phi$ as well. However, being a scalar, the perturbation of $\Phi$ will be exclusively of polar type and hence will not affect our computations involving axial perturbations alone. In addition, the perturbation of the energy momentum tensor of the fluid on the brane can be ignored, since that term is already multiplied by the four-dimensional Newton's constant and hence the contribution from it will be of higher order. Thus it follows that we may work in the regime where the energy-momentum tensor of the brane fluid can be neglected in comparison to the energy-momentum tensor of the radionlike field, i.e.,  $(\kappa^{2}/\ell)\delta T^{B}_{\mu\nu}\ll \delta T^{\phi}_{\mu\nu}$. Thus the evolution equation for the axial gravitational perturbation is given by
\begin{align}\label{gravitational-1}
\delta G_{\mu\nu}[h_{\alpha \beta}^{\rm axial}]=\frac{\delta T^{\phi}_{\mu\nu}[h_{\alpha \beta}^{\rm axial}]}{\Phi}~.
\end{align}
Further, in the present scenario involving static and spherically symmetric wormhole background, satisfying $R=0$, the above propagation equation reduces to the following form,
\begin{align}\label{grav-2}
\delta R_{\mu\nu}[h_{\alpha \beta}^{\rm axial}]=\frac{\delta T^{\phi}_{\mu\nu}[h_{\alpha \beta}^{\rm axial}]}{\Phi}~.
\end{align}
Using the gauge freedom to choose the Regge-Wheeler gauge, in which among the three axial metric perturbation components, only $h_{r\phi}$ and $h_{t\phi}$ survive, while $h_{\theta\phi}$ can be chosen to be zero \cite{PhysRev.108.1063}. Further, since the background geometry is static and spherically symmetric, one can decompose these two nonzero axial metric perturbations in the following manner
\begin{align}
h_{t\phi}=e^{-i\omega t} h_{0}(r)\sin\theta \partial_{\theta}P_{l}(\cos\theta)~,\\
h_{r\phi}=e^{-i\omega t} h_{1}(r)\sin\theta \partial_{\theta}P_{l}(\cos\theta)~.
\end{align} 
These two nonvanishing components of the metric perturbation induces the following three nonvanishing components of the perturbed Ricci tensor, namely, $\delta R_{t\phi}$, $\delta R_{r\phi}$ and $\delta R_{\theta\phi}$. The $\delta R_{\theta \phi}$ component of the perturbed Ricci tensor, when substituted in \ref{grav-2}, governing the evolution equation for the axial perturbation, with a background metric $g^{0}_{\mu\nu}=\textrm{diag.}(-f(r),(1/g(r)),r^{2},r^{2}\sin^{2}\theta)$, yields the following equation
\begin{align}\label{h-0-1}
h_{0}=\frac{i}{\omega}f(r)g(r)h_{1}~\frac{d}{dr}\left\{\ln \Phi+\ln[h_{1}\sqrt{f(r)g(r)}]\right\}~.
\end{align}
Note that this equation provides one of the perturbation variables, $h_{0}$ in terms of the other perturbation variable $h_{1}$ and the background scalar field $\Phi(r)$. Similarly, from the $(r,\phi)$ component of \ref{grav-2} we obtain 
\begin{align}\label{h-0-2}
i\omega r^{2}\frac{d}{dr}\left(\frac{h_{0}}{r^{2}}\right)=h_{1}\left[\omega^{2}+\frac{f(r)g(r)}{r}\frac{d\ln\Delta}{dr}-\frac{l(l+1)}{r^{2}}f(r)+f(r)g(r)H(\Phi)\right]~,
\end{align}
where we have defined $\Delta\equiv r^{2}f(r)g(r)$ and $H(\Phi)\equiv (1/\Phi)\left[(2/r)(d\Phi/dr)+\{1/2(1+\Phi)\}(d\Phi/dr)^{2}\right]$. Now we can substitute for $h_{0}$ from \ref{h-0-1} in \ref{h-0-2}, to get a single differential equation for the radial perturbation variable $h_{1}(r)$. In order to bring the resulting equation into the form of a time independent Schr\"odinger equation, we define the following master variable: $\psi^{(2)}_{lm}=(\sqrt{f(r)g(r)\Phi(r)}/r)h_{1}$. In terms of this master variable, we have our desired Regge-Wheeler-like equation for the background wormhole spacetime as
\begin{align}
\frac{d^{2}\psi^{(2)}_{lm}}{dr_{*}^{2}}+\left[\omega^{2}-V^{(2)}_{l}(r)\right]\psi^{(2)}_{lm}=0~,
\end{align}
where, $V_{l}^{(2)}(r)=V_{\rm g}(r)+V_{\Phi}(r)$, with $V_{\rm g}(r)$ is purely constructed out of background metric components and $V_{\Phi}(r)$ depends on the background radion like field $\Phi$. Below we provide the explicit formulas for these two potentials,
\begin{align}
V_{\rm g}(r)&=f(r)\left[\frac{l(l+1)}{r^{2}}-r\sqrt{\frac{g(r)}{f(r)}}\frac{d}{dr}\left(\frac{\sqrt{f(r)g(r)}}{r^{2}}\right)-\frac{g(r)}{r}\frac{d}{dr}(\ln\Delta)\right]~,
\label{V-s}
\\
V_{\Phi}(r)&=\frac{3}{4}\frac{f(r)g(r)}{\Phi^{2}}\frac{\left(\frac{d\Phi}{dr}\right)^{2}}{1+\Phi}~.
\label{V-Phi}
\end{align}
One can observe that if we had considered gravitational perturbations of any static and spherically symmetric four dimensional spacetime, with $R=0$, in the context of general relativity and had ignored the perturbations arising from the matter energy momentum tensor (as is the case here), then only the potential $V_{\rm g}(r)$ would be present. The contribution in the form of $V_{\Phi}$ solely arises from the existence of higher spacetime dimensions and while deriving the same, we have used the equation of motion for $\Phi(r)$ from \ref{EOM-Phi}, which can be expressed as
\begin{align}\label{Phi-EOM-tr}
\frac{d^{2}\Phi}{dr^{2}}+\frac{d \ln[r^{2}\sqrt{f(r)g(r)}]}{dr}\frac{d\Phi}{dr}=\frac{1}{2(1+\Phi)}\left(\frac{d\Phi}{dr}\right)^{2}~.
\end{align}
The solution of the above equation yields the value of the background radion like field which can be written as follows,
\begin{equation}
\Phi(r)=\frac{1}{4} \Phi _1^2 \{\log (f(r))\}^{2}+\sqrt{\Phi _0+1} \Phi _1 \log (f(r))+\Phi _0
\end{equation}
where $\Phi_0$ and $\Phi_1$ are constant of integration. These two parameter sets the interbrane separation $d=\frac{1}{2} \log (\Phi(r)+1)$. Now the distinction between the braneworld scenario and general relativity can be made clearer. In general relativity, no hair theorems \cite{PhysRevD.5.1239,PhysRevD.51.R6608} simply rule out any nontrivial scalar configuration outside the black hole, i.e., $\Phi$ becomes constant throughout the spacetime and we can see that it is consistent with \ref{Phi-EOM-tr}. In this case, $V_{\Phi}$ vanishes and the only contribution comes from $V_{\rm g}(r)$ alone. This also demands $p=0$ in the original wormhole spacetime, which yields, $-g_{tt}=[1-(2M/r)]=g^{rr}$ and hence we will get back the Regge-Wheeler equation for axial perturbation of the Schwarzschild spacetime. It is also instructive to note that among the above three kinds of perturbations-scalar, electromagnetic, and gravitational-only in the case of gravitational perturbation, the radion field makes its distinct contribution through the potential term. This is consistent with the fact that only gravity can sense the presence of extra dimensions, while other fields only sense the extra dimension through the metric alone. We will now try to solve these equations analytically using certain approximation methods and then shall use numerical techniques to obtain the quasinormal modes and the ringdown waveform of the braneworld wormhole, under all the three perturbations.  

\section{The spectrum of the quasinormal modes}\label{wormhole_qnm}

Now we are in a place to study the quasinormal mode (QNM) spectrum of the system. There are several approximate analytical methods as well as numerical techniques available for this purpose\cite{Konoplya:2011qq,Leaver:1985ax,PhysRevD.35.3621}. However, most of the analytical techniques do not work in the case of wormhole geometry. The WKB approximation, e.g., applies to black holes, since the potential involves one single maxima. However, in the present context, the potential involves two maxima and one minima, in which case it is not clear how the WKB method can be applied. The method of continued fractions also can not be applied, since this method only works for spacetimes with polynomial metric elements, while most of the wormhole spacetimes, including braneworld wormhole, involve square roots. Finally, the matching of asymptotic and near-the-throat solutions, yielding QNMs, is also not applicable in the present context, as the presence of the square root factor modifies the matching procedure significantly. Thus we will use a completely different method used for wormhole spacetimes, namely the transfer matrix method  \cite{Bueno:2017hyj}. This is what we elaborate below. 

\subsection{Approximate analytical methods}

As mentioned earlier, it is useful to present the wormhole spacetime as a black hole mimicker, which connects two identical spacetimes at the throat. Thus the wormhole spacetime can be described by a superposition of the potentials associated with these two individual spacetimes. Suppose, $V^{(s)}_{l}(r)$ denotes the potential, a spin $s$ perturbation experiences in one universe, then the spin $s$ perturbation in the wormhole spacetime will experience the following double potential \cite{Bueno:2017hyj},
\begin{align}\label{Model-V-l}
V^{(s)}_{l~\textrm{(wormhole)}}(r_{*})=\theta(r_{*})V_{l}^{(s)}\left(r_{*}-\frac{L}{2}\right)+\theta(-r_{*})V_{l}^{(s)}\left(-r_{*}-\frac{L}{2}\right)~.
\end{align} 
Note that $V_{l}^{(s)}(r_{*})$ appears in the following differential equation,
\begin{align}\label{Radial-black}
\frac{d^{2}\psi_{lm}^{(s)}}{dr_{*}^{2}}+[\omega^{2}-V_{l}^{(s)}(r_{*})]\psi_{lm}^{(s)}=0~,
\end{align}  
where, $\psi_{lm}^{(s)}$ represents the master radial perturbation variable, if we had considered one of the spacetime alone. Here, $s=0$, $s=1$, and $s=2$, corresponds to scalar, electromagnetic and axial gravitational perturbations, respectively. Using the asymptotic behavior of the potential $V_{l}^{(s)}(r_{*})$ near infinity and in near-the-throat region, for various spins of the perturbation, we obtain the following behavior for the master perturbation variable\footnote{In this context, one should keep in mind that for black holes, the limit $r_{*}\rightarrow -\infty$ corresponds to the near-horizon regime, but in the case of wormholes, $r_{*}\rightarrow 0$ corresponds to the near-the-throat region.},
\begin{align}\label{psi-u-nature} 
\psi_{lm}^{(s)}(r_{*})=
\begin{cases} 
Ae^{-i\omega r_{*}}+Be^{i\omega r_{*}} & \text{for}\ \quad r_{*}\to \infty
\\
Ce^{-i(\omega-\omega_{0}^{(s)})r_{*}}+De^{i(\omega-\omega_{0}^{(s)})r_{*}} & \text{for} \quad r_{*}\to 0~,
\end{cases} 
\end{align}
here $\omega_{0}^{(s)}$ is defined by the relation $\sqrt{\omega^{2}-V_{l}^{(s)}(r_{*}\rightarrow 0)}\equiv (\omega-\omega_{0}^{(s)})$. Now we introduce the transfer matrix $T$ associated with this scattering problem. The job of this matrix is to relate the ingoing and outgoing amplitudes through the potential $V_{l}^{(s)}$, at one side of the wormhole alone. In the present context, the transfer matrix for the Universe with $r_{*}>0$, takes the following form:\\
\begin{align}\label{transfer}
\begin{pmatrix}
B\\
A
\end{pmatrix}=T\begin{pmatrix}
D\\
C
\end{pmatrix}~.
\end{align}
Use of \ref{Model-V-l} and \ref{transfer} depicts that, the transfer matrix $\mathbb{T}$ for the scattering problem through the full wormhole potential $V_{l~\textrm{wormhole}}^{(s)}$ requires transfer matrices, through the individual potentials $V_{l}^{(s)}(r_{*})$ and $V_{l}^{(s)}(-r_{*})$, along with propagation through the throat. This is obtained by matching the solution for $V_{l}^{(s)}(-r_{*})$ with $V_{l}^{(s)}(r_{*})$ in the region between the two potentials, yielding \cite{Bueno:2017hyj},
\begin{align}\label{tranfer-V}
\mathbb{T}=T\begin{pmatrix}
e^{i(\omega-\omega_{0})L}&0\\
0&e^{-i(\omega-\omega_{0})L}
\end{pmatrix}\sigma_{x}T^{-1}\sigma_{x}~,
\end{align}
where, $\sigma_{x}$ is the first one, among the three Pauli $\sigma$ matrices. The QNMs, associated with the wormhole spacetime is characterized by the boundary condition that, there is no ingoing wave from $r_{*}=\pm \infty$. To manifest this condition on the universe with $r_{*}>0$, we set $A=0$ in \ref{psi-u-nature}, which gives
\begin{align}
\frac{D}{C}\equiv R^{(s)}(\omega)=-\frac{T_{21}}{T_{22}}~,
\end{align}
where, $R^{(s)}(\omega)$ corresponds to the reflectivity of the potential experienced by spin s perturbation in both the universes, connected by the wormhole throat. Similarly manifesting an identical condition in the universe with $r_{*}<0$, we obtain the condition $\mathbb{T}_{22}=0$ for the wormhole spacetime and it yields the spectrum of the QNM frequencies, in terms of the reflectivity of the photon sphere as,
\begin{align}\label{Wormhole-QNF}
e^{-i(\omega_{n}-\omega_{0})L}=-e^{-in\pi}R^{(s)}(\omega_{n})~,
\end{align}
where, $\omega_{n}$ corresponds to the QNM frequency corresponding to the $n$th overtone. Finally, the master perturbation variable for the wormhole spacetime, associated with the $n$th order QNM, can be expressed as 
\begin{align}\label{Wormhole-QNM}
\psi^{(s)}_{lm~\textrm{(wormhole)}}(r_{*},\omega_{n})=\theta(r_{*})\psi^{(s)}_{lm}\left(r_{*}-\frac{L}{2}\right)-e^{-in\pi}(r_{*}\rightarrow -r_{*})~,
\end{align}
where $\psi_{lm}^{(s)}$ is the solution of \ref{Radial-black} with the boundary condition that there are no ingoing waves at $r_{*}\rightarrow \infty$. 

Therefore, the QNM frequencies depend crucially on the reflectivity of the photon sphere in one of the universes forming the wormhole spacetime. However, the presence of the square root in the metric elements prohibits us in order to solve for the above reflectivity in an exact manner. Rather, we are forced to certain approximation methods in order to derive the QNM frequencies analytically. This is achieved by approximating the potentials $V_{l}^{(s)}$ by the P\"oschl-Teller potential, 
\begin{align}\label{Posch-poten}
V^{\rm PT}(r_{*})=V_{0} \sech^{2}\left[\alpha_{0} \left(r_{*}-r^{\rm max}_{*}\right)\right]~,
\end{align} 
where, $V_{0}$ and $\alpha_{0}$ are parameters, and $r_{*}^{\rm max}$ corresponds to the maxima of the potential associated with the photon sphere in both the universes, on either side of the throat. The wormhole potential $V_{l~\textrm{(wormhole)}}^{(s)}$, on the other hand, corresponds to a double P\"oschl-Teller potential. It is known \cite{Konoplya:2011qq} that the P\"oschl-Teller potential gives correct black hole QNMs for large angular momentum, that is for $l\gg 1$ and gives only $2\%$ relative error for $l=2$. This information may be taken as the motivation for using it as a model for $V_{l}^{(s)}$. To ensure that the double P\"oschl-Teller Potential constructed using the prescription in \ref{Model-V-l}, mimic the true effective potential $V_{l~\textrm{(wormhole)}}^{(s)}$ for the wormhole spacetime, we need to fix the parameters $V_{0}$ and $\alpha_{0}$ in such a way that
\begin{align}
V_{0}&=V_{l}^{(s)}(r^{\rm max}_{*})~,
\\
\alpha^{2}_{0}&=-\frac{1}{2V_{0}}\frac{d^{2}V_{l}^{(s)}}{dr^{2}_{*}}\bigg|_{r_{*}=r_{*}^{\rm max}}~.
\end{align} 
In the present scenario, we can calculate the values of these parameters for the most simplest case of scalar perturbation (that is for $V_{l}^{(0)}(r(r_{*}))$~, along with that of $r_{*}^{\rm max}$, by using the following approximate analytic expressions (with $l\neq 0$):
\begin{align}
r_{\rm max}&\simeq\frac{3M}{2(1+p)}\left(1-\frac{1}{\eta^{2}}\right)\left(1+\sqrt{1+\frac{32\eta^{2}}{9(\eta^{2}-1)^{2}}}\right)~,
\\
V_{0}&\simeq \frac{(1+p)^{2}\eta^{2}}{27M^{2}}+\frac{2(1+p)^{2}}{81M^{2}}~,
\\
\alpha^{2}_{0}&\simeq \left[\frac{(1+p)y(r_{\rm max})}{9M}\right]^{2}\frac{(2\eta^{2})^{5}}{(\eta^{2}-1)^{4}}\frac{1}{\frac{\eta^{2}}{27}+  \frac{2}{81}}\frac{\sqrt{1+\frac{32\eta^{2}}{9(\eta^{2}-1)^{2}}}}{\left[1+\sqrt{1+\frac{32\eta^{2}}{9(\eta^{2}-1)^{2}}}\right]^{5}}~,
\end{align}
where, $\eta^{2}\equiv l(1+1)$ and in deriving these formulas we have assumed $p\ll 1$. The next task is to compute the reflectivity of the P\"oschl-Teller potential, which can be performed using the standard matching procedure in quantum mechanics \cite{poschl1933bemerkungen} and the reflectivity takes the following form,
\begin{align}\label{Reflectivity of the poschl}
R_{\rm PT}(\omega)=-\frac{\Gamma(1+i\frac{\omega}{\alpha_{0}})\Gamma(\xi-i\frac{\omega}{\alpha_{0}})\Gamma(1-\xi-i\frac{\omega}{\alpha_{0}})}{\Gamma(1-i\frac{\omega}{\alpha_{0}})\Gamma(\xi)\Gamma(1-\xi)}~;\qquad \xi \equiv \frac{1\pm i\sqrt{\frac{4V_{0}}{\alpha^{2}_{0}}-1}}{2}
\end{align}
One can hence derive the QNM frequencies, using the reflectivity derived above and following \ref{Wormhole-QNF}. A more compact expression of the QNM frequencies can be obtained in the region, where $(V_{0}/\alpha^{2})\rightarrow 0$, i.e., the strength of the potential is small, but its characteristic length scale is large. In this limit, one obtains \cite{Bueno:2017hyj}
\begin{align}\label{QNM-double-dirac}
\omega_{n}&\approx \frac{n\pi}{L}\left[\left(1-\frac{2}{L\nu}\right)-i\frac{2n\pi}{L^{2}\nu^{2}}+...\right]~\text{with}\quad\nu\equiv\frac{2V_{0}}{\alpha_{0}}~.
\end{align}
Using these results in the present context, we can obtain analytic expressions for the QNM frequencies, some of which have been listed in \ref{tab1}.

\begin{table}[th!]
	\centering
		\def\arraystretch{1.3}
	\setlength{\tabcolsep}{1.5em}
	\begin{tabular}{|p{1cm}||p{5cm}|p{5cm}|  }
	\hline
	Mode $ n $    & $M\omega_n~(l=1)$                       & $M\omega_n~(l=2)$                           \\ \hline
		$ 1  $    & $0.01723-3.520.10^{-6}i $ & $0.01735-5.307.10^{-7}i $   \\ \hline
		$ 2   $   & $0.03447-1.408.10^{-5}i$   & $0.03471-2.123.10^{-6}i$  \\ \hline
		$ 3     $ & $0.05171-3.168.10^{-5}i $  & $0.05207-4.777.10^{-6} $ \\ \hline
		$ 4    $  & $0.06894-5.633.10^{-5}i$   & $0.06943-8.492.10^{-6}i$  \\ \hline
		$ 5     $ & $0.08618-8.801. 10^{-5}i $  & $0.08678-1.327.10^{-5}i $ \\ \hline
		$ 6    $  & $0.1034-1.267.10^{-4}i $  & $0.1041-1.911.10^{-5} $ \\ \hline
    	$ 7     $ & $0.1206-1.725.10^{-4}i$   & $0.1215-2.601.10^{-5}i$  \\ \hline
		$ 8   $   & $0.1378-2.253.10^{-4}i $  & $0.1388-3.397.10^{-5}i$   \\ \hline
		$ 9    $  & $0.1551-2.852.10^{-4}i $    & $0.1562-4.299.10^{-5} $   \\ \hline
		$ 10  $   & $0.1723-3.520.10^{-4}i $    & $0.1735-5.307.10^{-5}i $   \\ \hline
	\end{tabular}
	\caption{Approximate analytical computation of the QNM frequencies, using the P\"oschl-Teller potential, for the massless scalar perturbation of the braneworld wormhole, with the braneworld parameter $p=10^{-10}$. These values must be contrasted with the corresponding ones obtained through numerical methods in the next section.}\label{tab1}
\end{table}

Note that the above only provides the QNM frequencies under various sets of approximations and thus it is worthwhile to consider the exact QNM frequencies by solving the perturbation equations for the master variables of scalar, electromagnetic and gravitational perturbations. This will also enable us to understand the ringdown waveform at late times and possible emergence of echoes. This is what we perform next.

\subsection{Numerical methods to obtain the quasinormal modes and the time-domain signal}

\begin{figure}
	\centering
	\minipage{0.48\textwidth}
	\includegraphics[width=\linewidth]{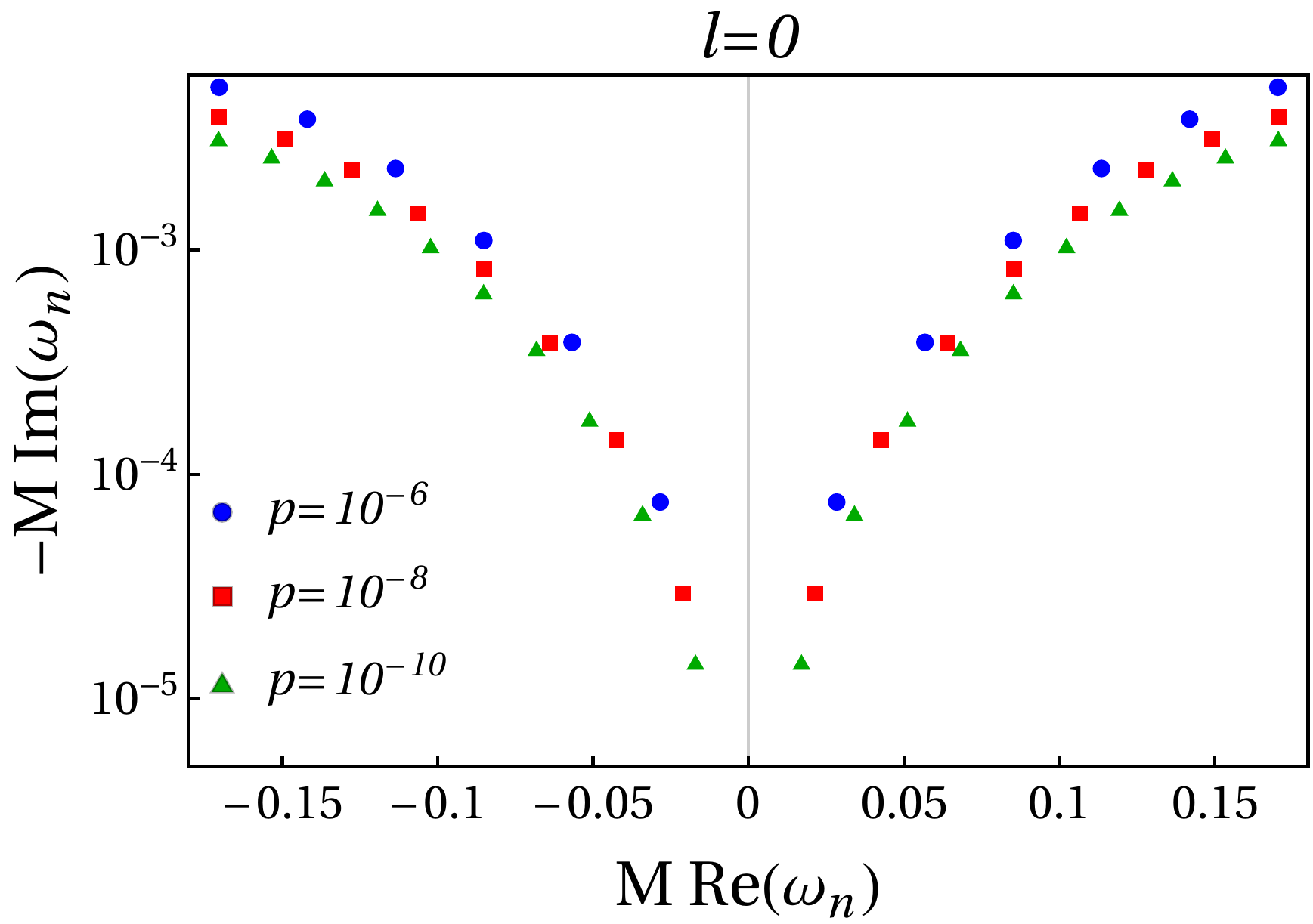}
	\endminipage\hfill
	\minipage{0.48\textwidth}
	\includegraphics[width=\linewidth]{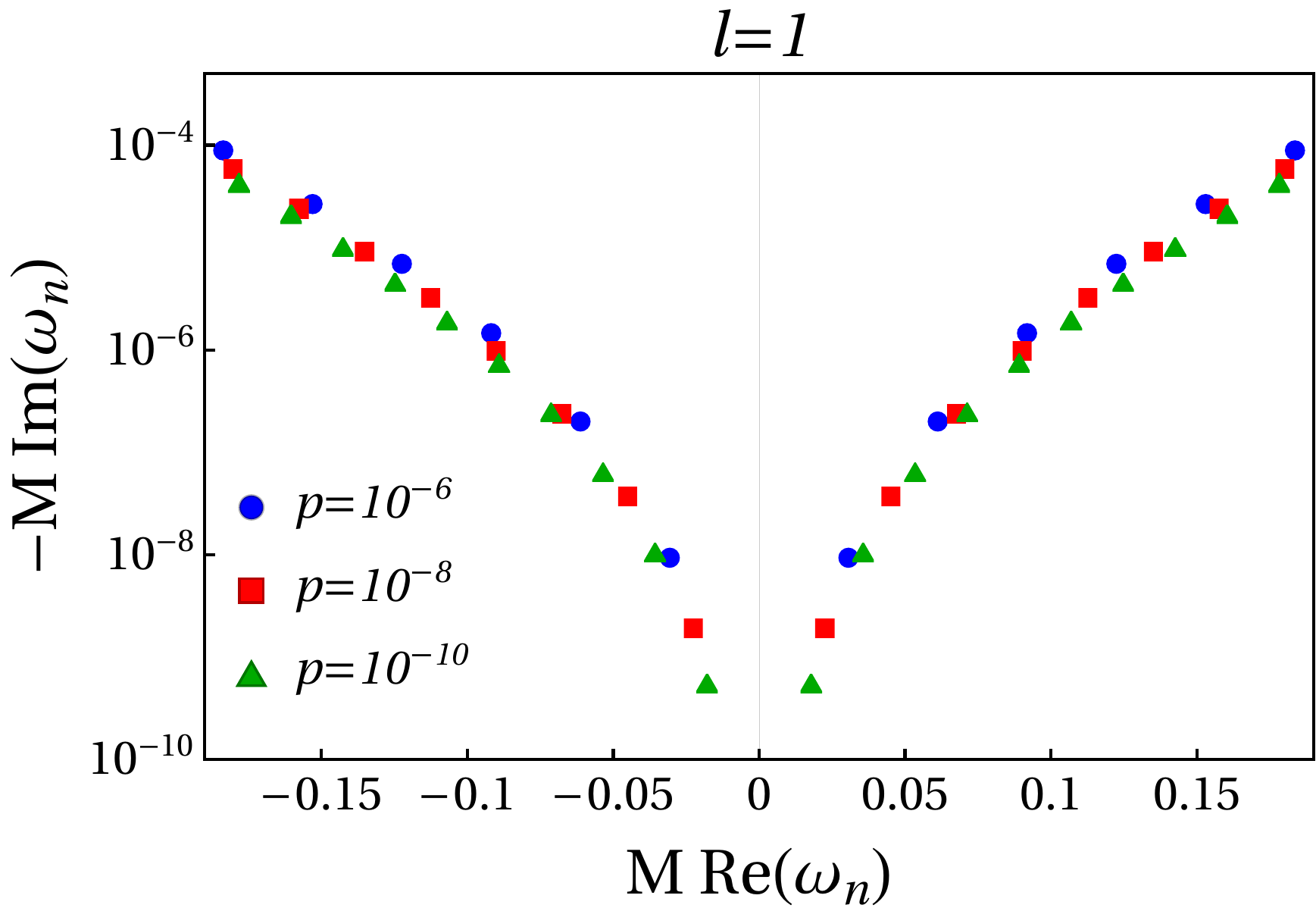}
	\endminipage
	\caption{The real and imaginary parts of the QNM frequencies, associated with scalar perturbation of the wormhole geometry have been presented for $l=0$ (left panel) and $l=1$ (right panel) and for different values of the wormhole parameter $p$.}\label{fig_QNM_Scalar}
\end{figure}

\begin{figure}
	\centering
	\minipage{0.48\textwidth}
	\includegraphics[width=\linewidth]{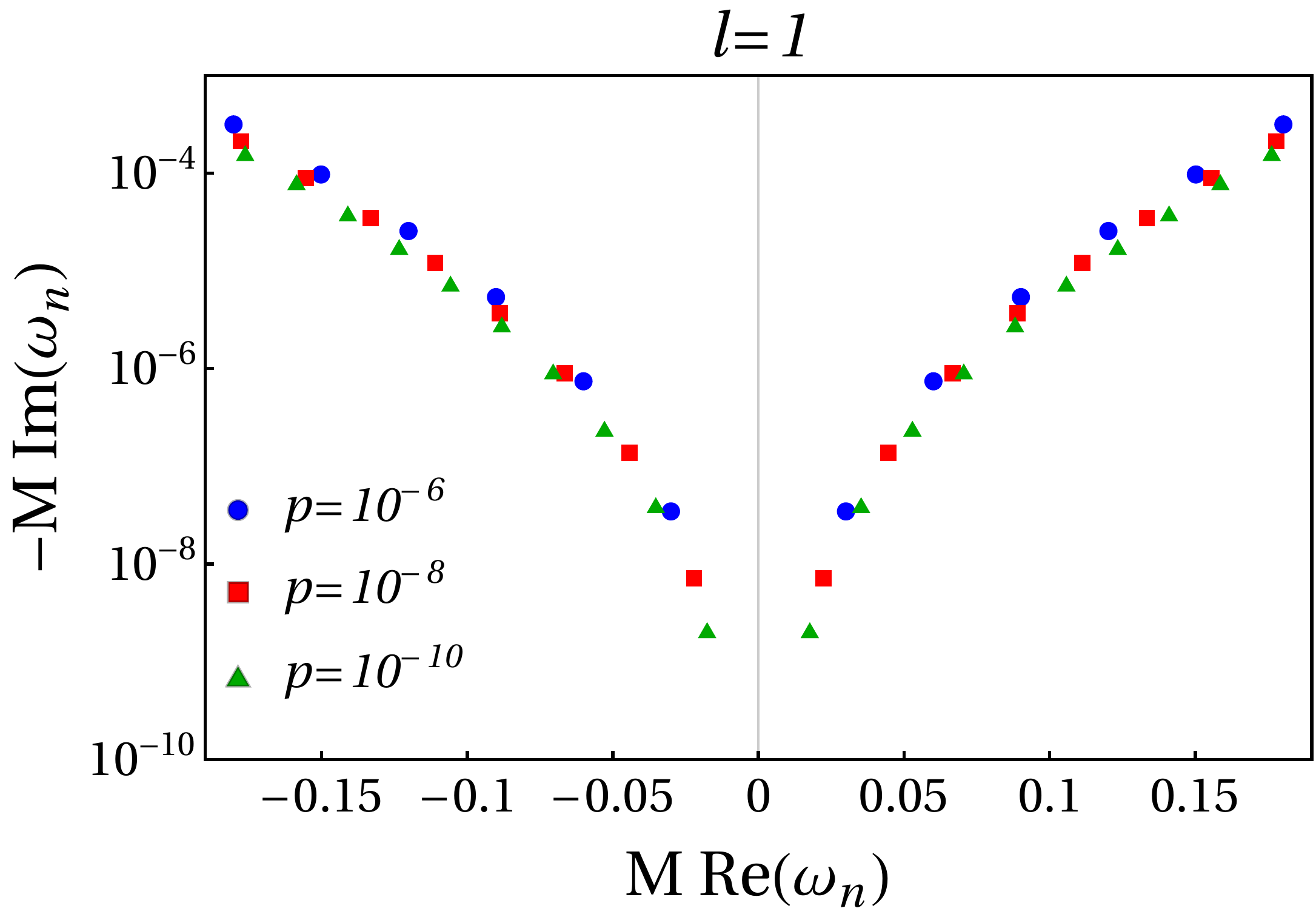}
	\endminipage\hfill
	\minipage{0.48\textwidth}
	\includegraphics[width=\linewidth]{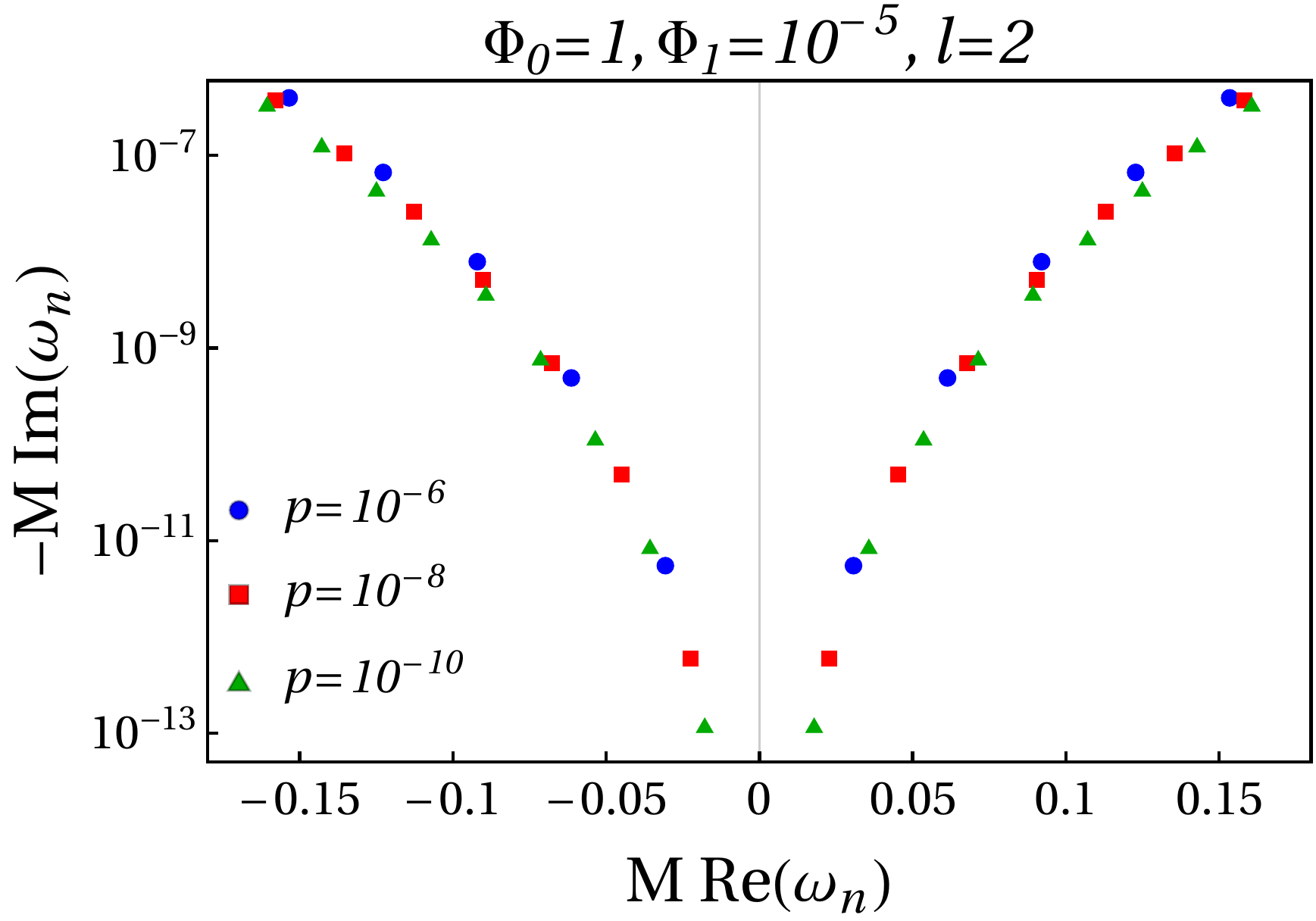}
	\endminipage
	\caption{The lowest-lying QNM frequencies of the wormhole under electromagnetic, the $l=1$ modes (left panel) and gravitational, the $l=2$ modes (right panel), perturbations have been presented. For the computation of the gravitational QNM frequencies, we consider the following choices for the radion-like field, $\Phi_0=1$ and $\Phi_1=10^{-5}$.}\label{fig_QNM_em}
\end{figure}

In this section, we briefly describe the numerical methods used to obtain the QNMs and the time-domain signal from the wormhole. As discussed earlier, the QNMs are the solutions of \ref{Radial-black} with $V_{l~\textrm{(wormhole)}}^{(s)}$ as the potential and outgoing boundary conditions at both the infinities i.e.,
\begin{equation}\label{b.c.}
\psi_{lm~\textrm{(wormhole)}}^{(s)}(r_*)\sim e^{\pm i\omega r_*},\qquad{r_*}\to\pm \infty~.
\end{equation}
In this paper, we make use of \ref{Wormhole-QNF} to find the quasinormal frequencies following the method prescribed in Ref.~\cite{Bueno:2017hyj}. One crucial part of the calculation is to obtain the expression of the reflection coefficient $R^{(s)}(\omega_{n})$. Note that, $\omega_{0}$ is a solution of \ref{Wormhole-QNF} for $R^{(s)}(\omega_{0})=-1$. This basically means that an incoming wave with frequency $\omega=\omega_{0}$ is fully reflected back to infinity by the potential barrier at the photon sphere. In other words, there is no wave inside the cavity formed by the double bump potential. Although the solution is unphysical, we can use it to obtain an approximate solution of \ref{Wormhole-QNF} around $\omega=\omega_{0}$. Furthermore,  we make use of the assumption that the width of the potential barrier is much smaller than the length of the cavity \cite{Bueno:2017hyj}. The assumption dictates that the real part of quasinormal frequency is of the order $\textrm{Re}(\omega_{n})\sim \omega_{0}+n\pi/L$ (see \ref{Wormhole-QNF}). This is because the width of potential governs the value of the reflection coefficient. We can see this more clearly from the analytical expression of the QNM frequencies presented in \ref{QNM-double-dirac}. The width of the P\"oschl-Teller potential is governed by the parameter $\alpha_{0}$; the width of the potential decreases with the increase of the parameter $\alpha_{0}$. Hence, the assumption $L\gg 1/\alpha_{0}$ leads to the conclusion $\textrm{Re}(\omega_{n})\sim n\pi/L$. Thus, in order to get an approximate value of the quasinormal frequencies, we expand $\omega_{n}$ and $R^{(s)}(\omega_{n})$ around $\omega_{0}$ in the following manner,
\begin{equation}\label{expand}
\begin{aligned}
\omega_{n}&=\omega_{0}+\sum_{k=1}^{\infty}\frac{C_k}{L^k}~,\\
R^{(s)}(\omega_{n})&=-1+\sum_{k=1}^{\infty}\frac{1}{k!}\frac{d^k R^{(s)}}{d\omega^k}\bigg|_{\omega=\omega_{0}}(\omega_{n}-\omega_{0})^k~.
\end{aligned}
\end{equation}
Replacing \ref{expand} in \ref{Wormhole-QNF}, we find the first three nonzero coefficients to be,
\begin{equation}\label{coefficients}
\begin{aligned}
C_1&=n\pi,\qquad{C_2}=-i n\pi R^{(s)'}(\omega_{0})\\
C_3&=-i\left[-in\pi\left(R^{(s)'}(\omega_{0})\right)+\frac{(n\pi)^2}{2}\left(R^{(s)''}(\omega_{0})+(R^{(s)'}(\omega_{0}))^2\right)\right]
\end{aligned}
\end{equation}
Using \ref{Wormhole-QNF} and \ref{coefficients}, we obtain the expression for the real part of the QNM frequencies as,
\begin{equation}\label{Re_QNM}
\begin{aligned}
\textrm{Re}(\omega_{n})&=\textrm{Re}\left[\left(\frac{n\pi}{L}-i\frac{n\pi}{L^2}R^{(s)'}(\omega_{0})\right)-i\left(\frac{(n\pi)^2}{2L^3}\left(R^{(s)''}(\omega_{0})+\left(R^{(s)'}(\omega_{0})\right)^2\right)\right)\right]~\\&=\frac{n\pi}{L}\left(1+\frac{1}{L}\textrm{Im}\left[R^{(s)'}(\omega_{0})\right]+\frac{n\pi}{2L^2}\left(\textrm{Im}\left[R^{(s)''}(\omega_{0})\right]+\textrm{Im}\left[R^{(s)'}(\omega_{0})\right]^2\right)\right)
\end{aligned}
\end{equation}
In a similar manner, we can obtain the imaginary part of the QNM frequencies by taking the absolute value of \ref{Wormhole-QNF}, which leads to the following expression
\begin{equation}\label{Im_QNM}
\textrm{Im}(\omega_{n})=\frac{1}{L}\textrm{log}[R^{(s)}(\omega_{n})]\approx\frac{\big|R^{(s)}\left(\textrm{Re}(\omega_{n})\right)\big|-1}{L}~.
\end{equation}
In the last step, we make use of the fact $\big|R^{(s)}(\omega_{n})\big|=1+\mathcal{O}(1/L)$. We obtain the reflection coefficient $R^{(s)}(\omega)$ by numerically solving \ref{Radial-black}, with the boundary condition presented in \ref{psi-u-nature}, and taking into consideration that there is no incoming wave from asymptotic infinity. Replacing the expression for $R^{(s)}(\omega)$ in \ref{Re_QNM} and \ref{Im_QNM}, by the one derived using numerical methods, we find the QNM frequencies of the wormhole spacetime. Following this analysis, in \ref{fig_QNM_Scalar}, we depict the QNM frequencies associated with the scalar perturbation for $l=0$ (left panel) and $l=1$ (right panel), and for different values of the wormhole parameter $p$. The lowest lying QNM frequencies for the electromagnetic ($l=1$ modes) and gravitational ($l=2$ modes) perturbations are shown in \ref{fig_QNM_em}. Note that, the computation of the gravitational QNMs requires the knowledge of the background radion-like field. Here, we choose the parameters as $\Phi_0=1$ and $\Phi_1=10^{-5}$. The parameters are so chosen that the interbrane distance $d=\log (\Phi(r)+1)/2$ remains non-negative. It is clear from the figures, that the imaginary parts of the QNM frequencies are very small ($\sim 10^{-4}-10^{-9}$) and this implies that, unlike black holes, the perturbations decay very slowly. Combining this fact with the reflectivity of the wormhole throat inevitably suggests the existence of echoes. In order to see these echoes explicitly, we perform the time-domain analysis below. 

\begin{figure}[!h]
	\minipage{0.33\textwidth}
	\includegraphics[width=\linewidth]{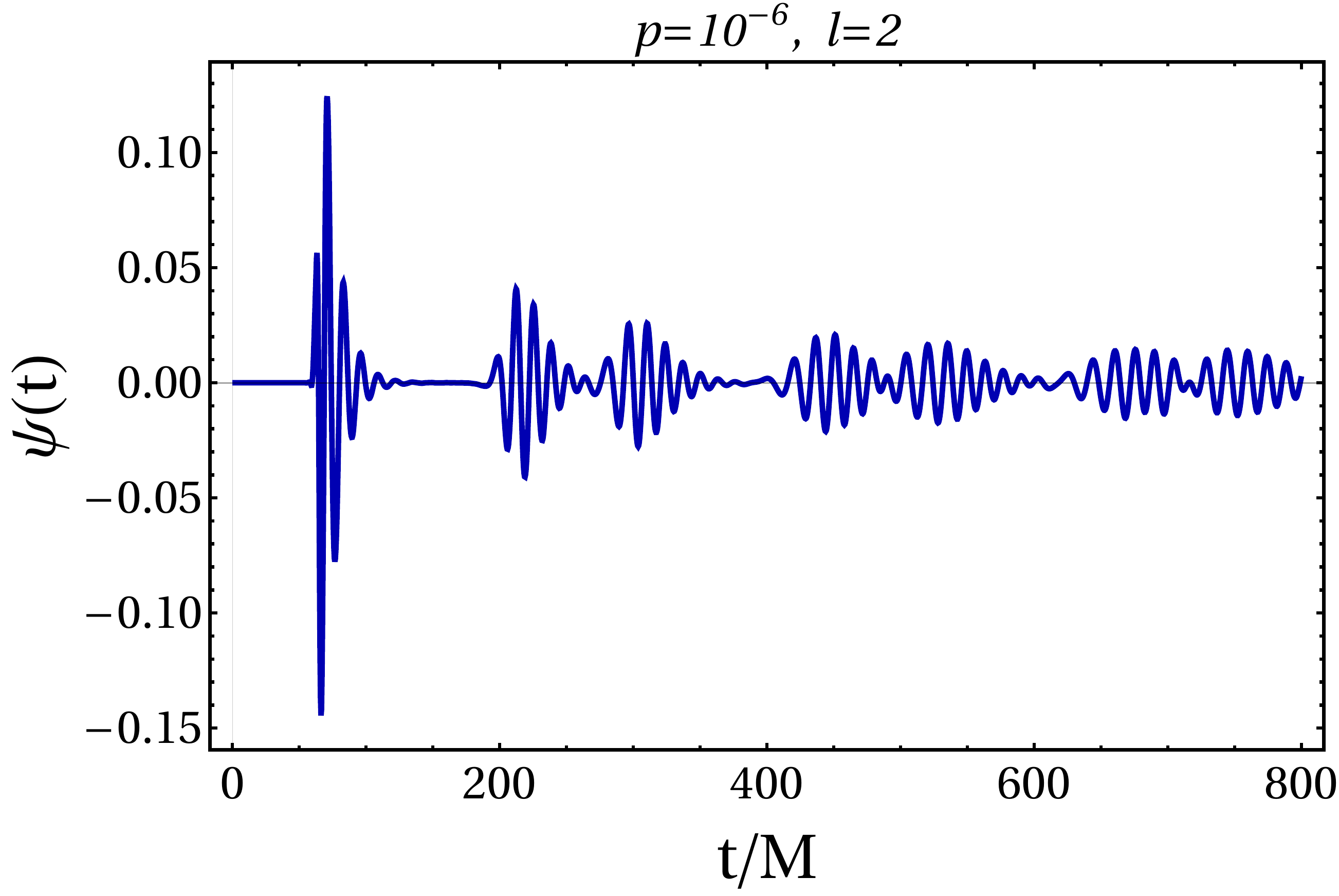}
	\endminipage\hfill
	\minipage{0.33\textwidth}
	\includegraphics[width=\linewidth]{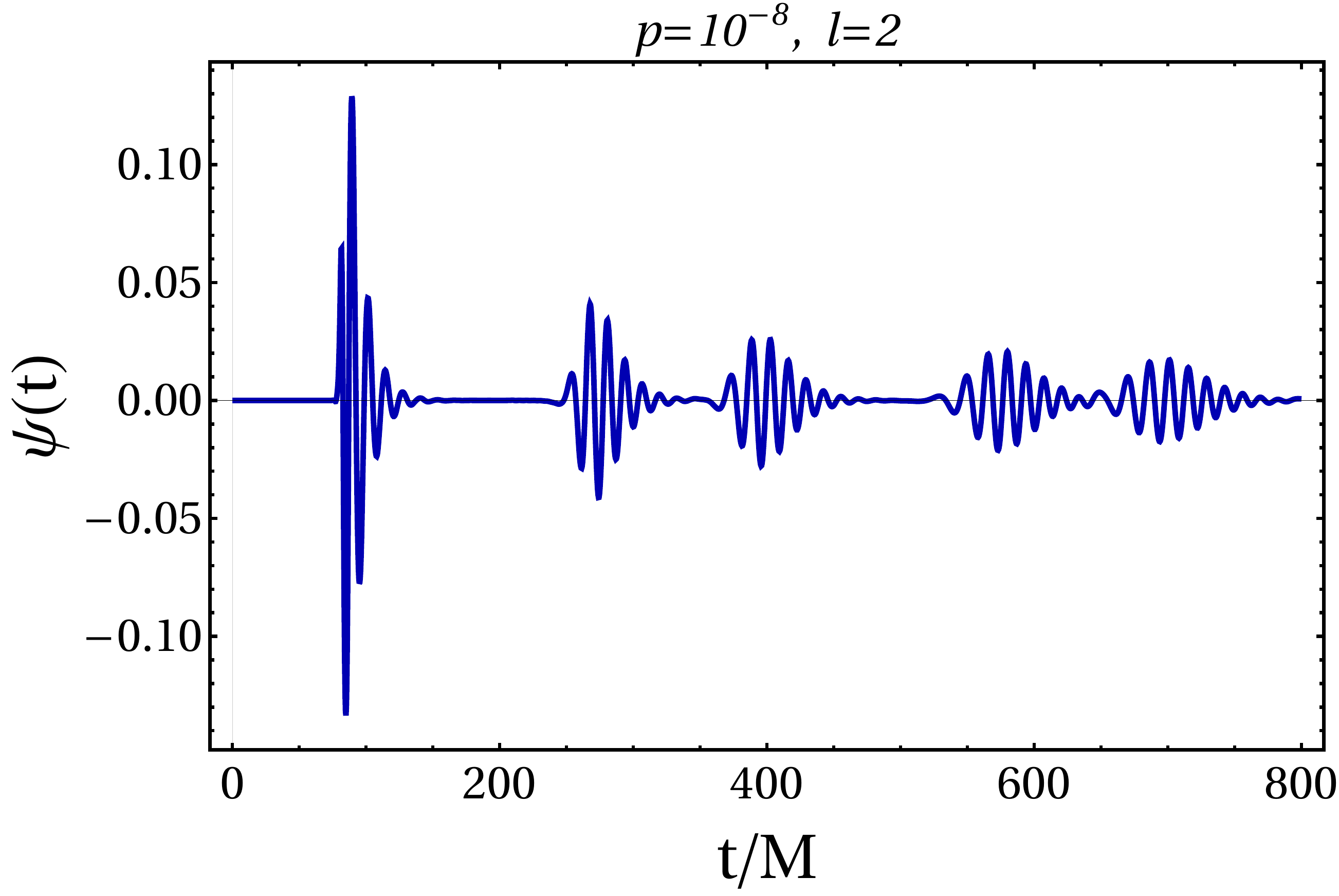}
	\endminipage\hfill
	\minipage{0.33\textwidth}%
	\includegraphics[width=\linewidth]{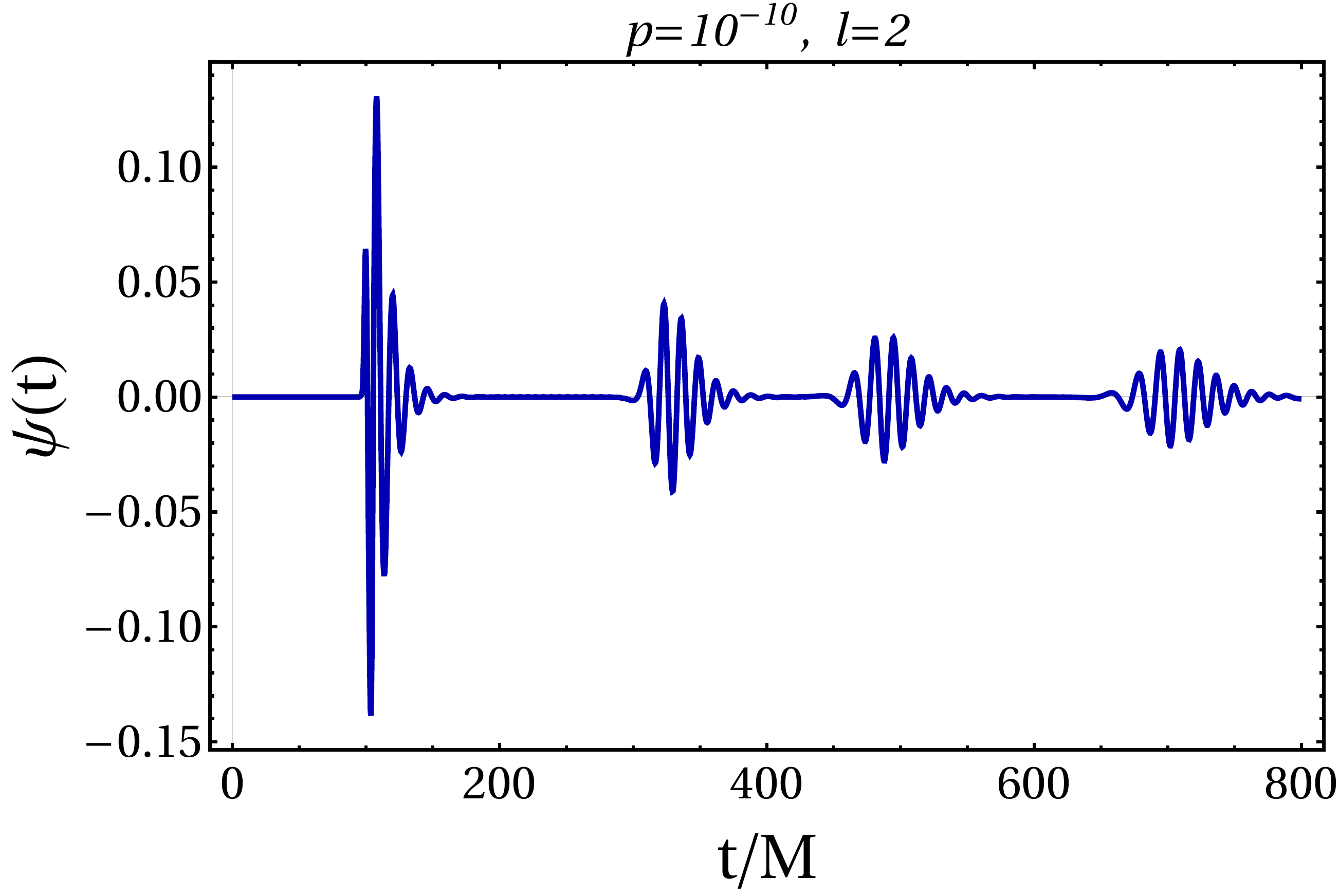}
	\endminipage
	\caption{The time-domain profiles of the scalar perturbation at $r=10M$ for $l=2$ and $p=10^{-6}$ (left panel), $p=10^{-8}$ (middle panel) and $p=10^{-10}$ (right panel), have been presented. Here, we consider a Gaussian initial profile $\hat{\psi}_{lm}(0,r_*)=0$, $\partial_{t}\hat{\psi}_{lm}(0,r_*)=e^{-(r_*-7)^2}~$ to solve the wavelike equation, presented in \ref{wave_eqn}. As evident, the time-separation between two consecutive echo signals increases with the decrease of the value of $p$.}\label{fig_ringdown_Scalar}
\end{figure}

\begin{figure}[!h]
	\minipage{0.33\textwidth}
	\includegraphics[width=\linewidth]{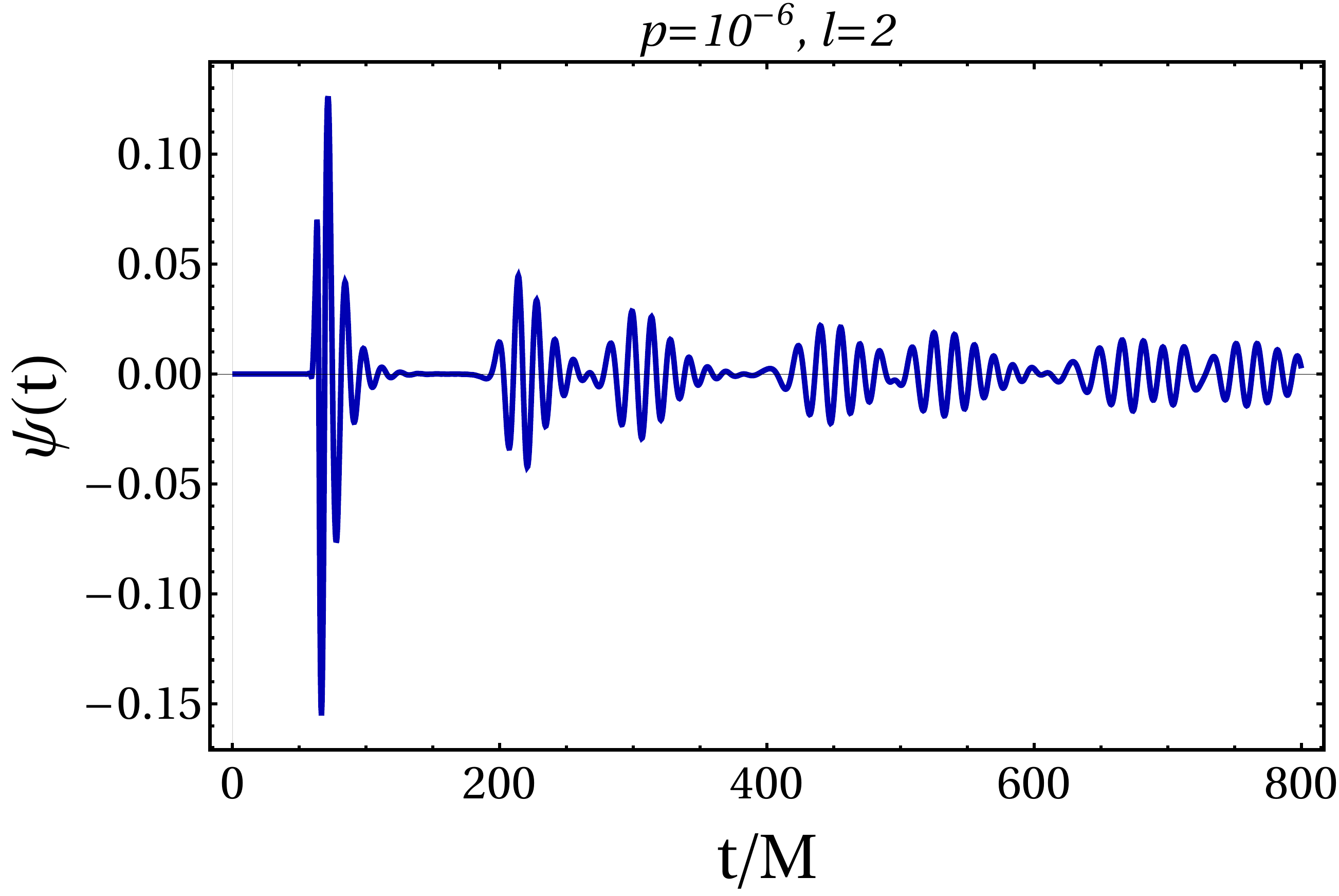}
	\endminipage\hfill
	\minipage{0.33\textwidth}
	\includegraphics[width=\linewidth]{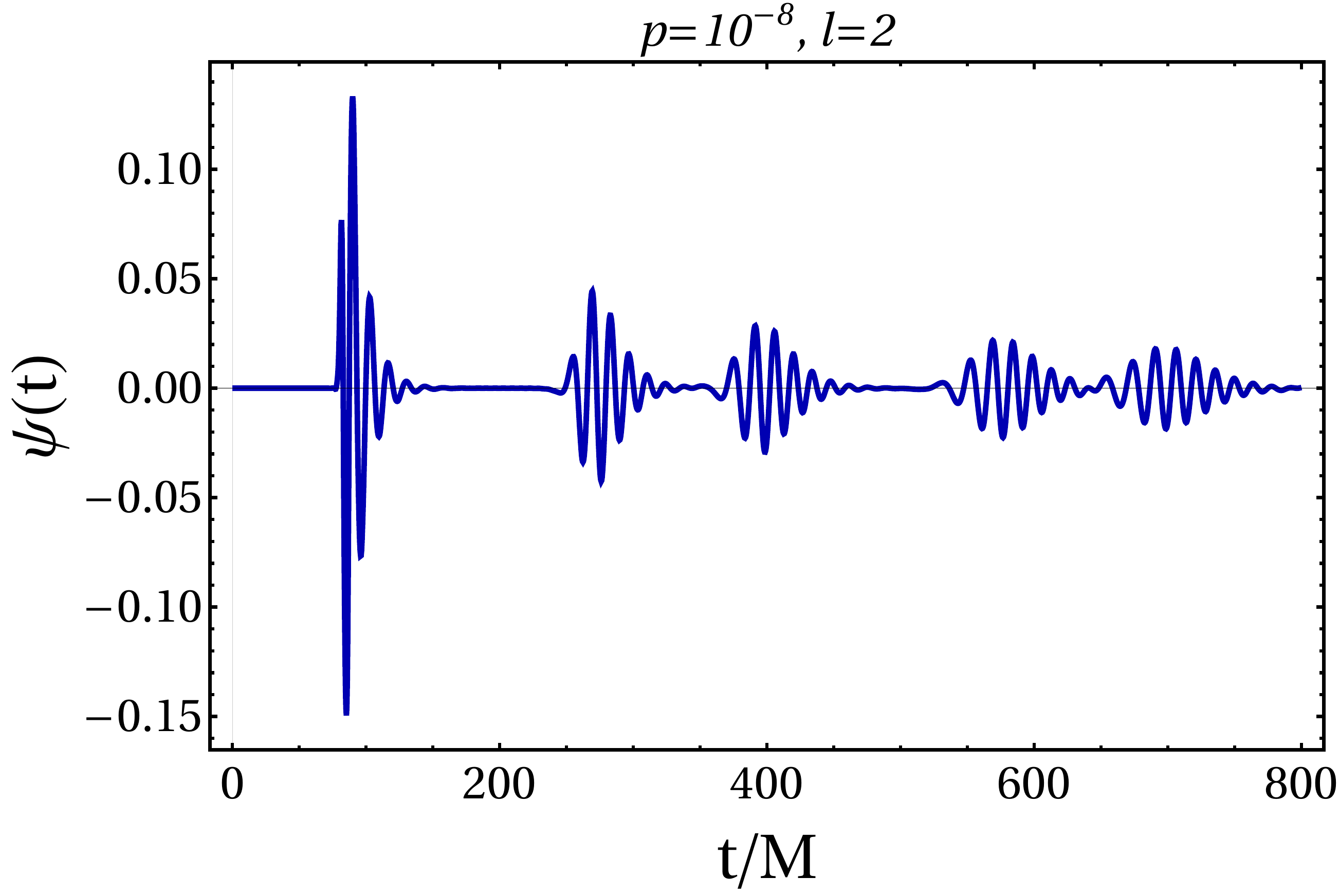}
	\endminipage\hfill
	\minipage{0.33\textwidth}%
	\includegraphics[width=\linewidth]{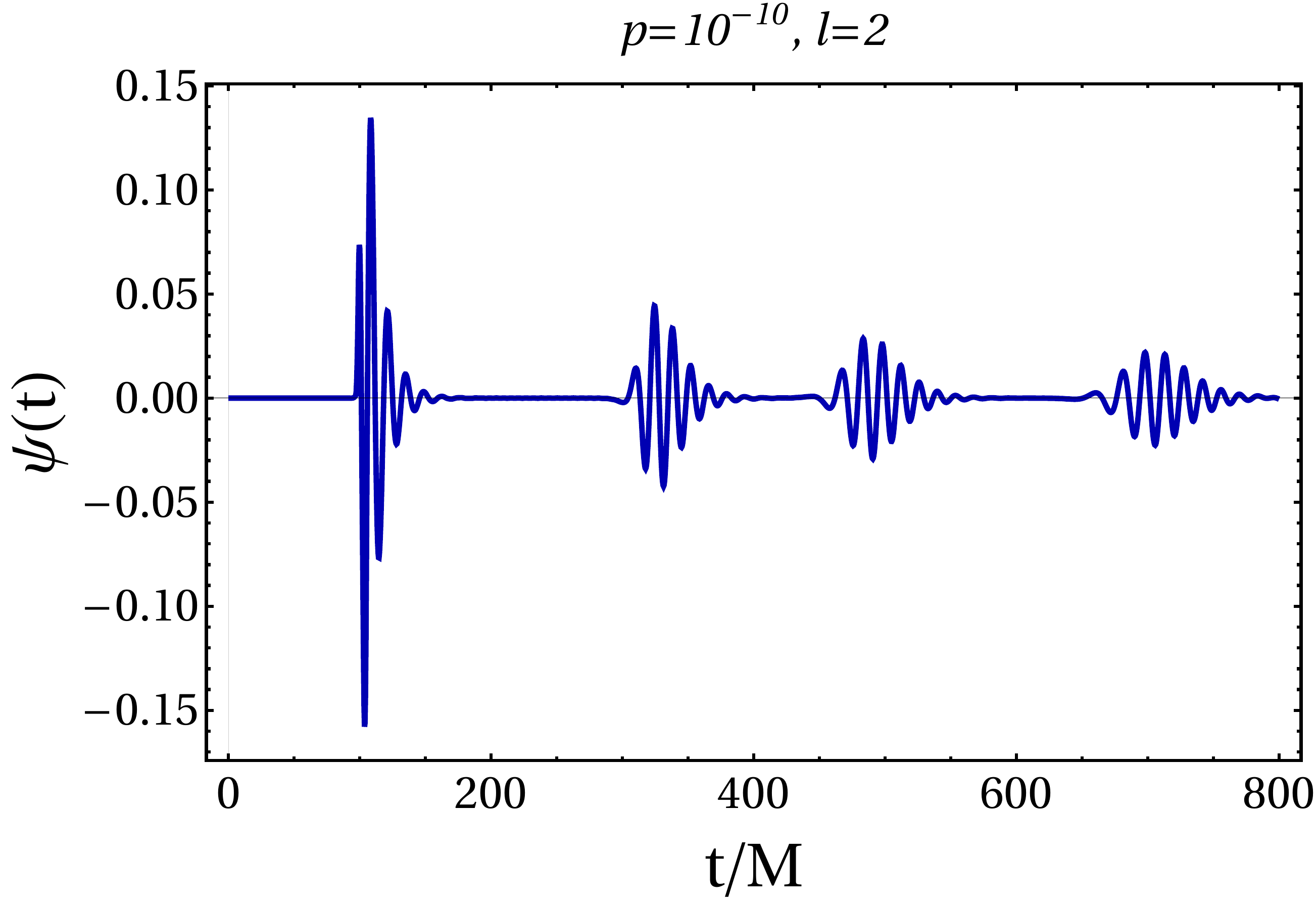}
	\endminipage
	\caption{The time-domain profiles of the electromagnetic perturbation at $r=10M$ for $p=10^{-6}$ (left panel), $p=10^{-8}$ (middle panel) and $p=10^{-10}$ (right panel), have been presented. We consider a Gaussian initial profile $\hat{\psi}_{lm}(0,r_*)=0$, $\partial_{t}\hat{\psi}_{lm}(0,r_*)=e^{-(r_*-7)^2}~$, identical to that of scalar perturbation, to solve the wavelike equation in \ref{wave_eqn}. Here also, with a decrease in $p$, the time gap between echoes increases.}\label{fig_ringdown_em}
\end{figure}

\begin{figure}
	\minipage{0.33\textwidth}
	\includegraphics[width=\linewidth]{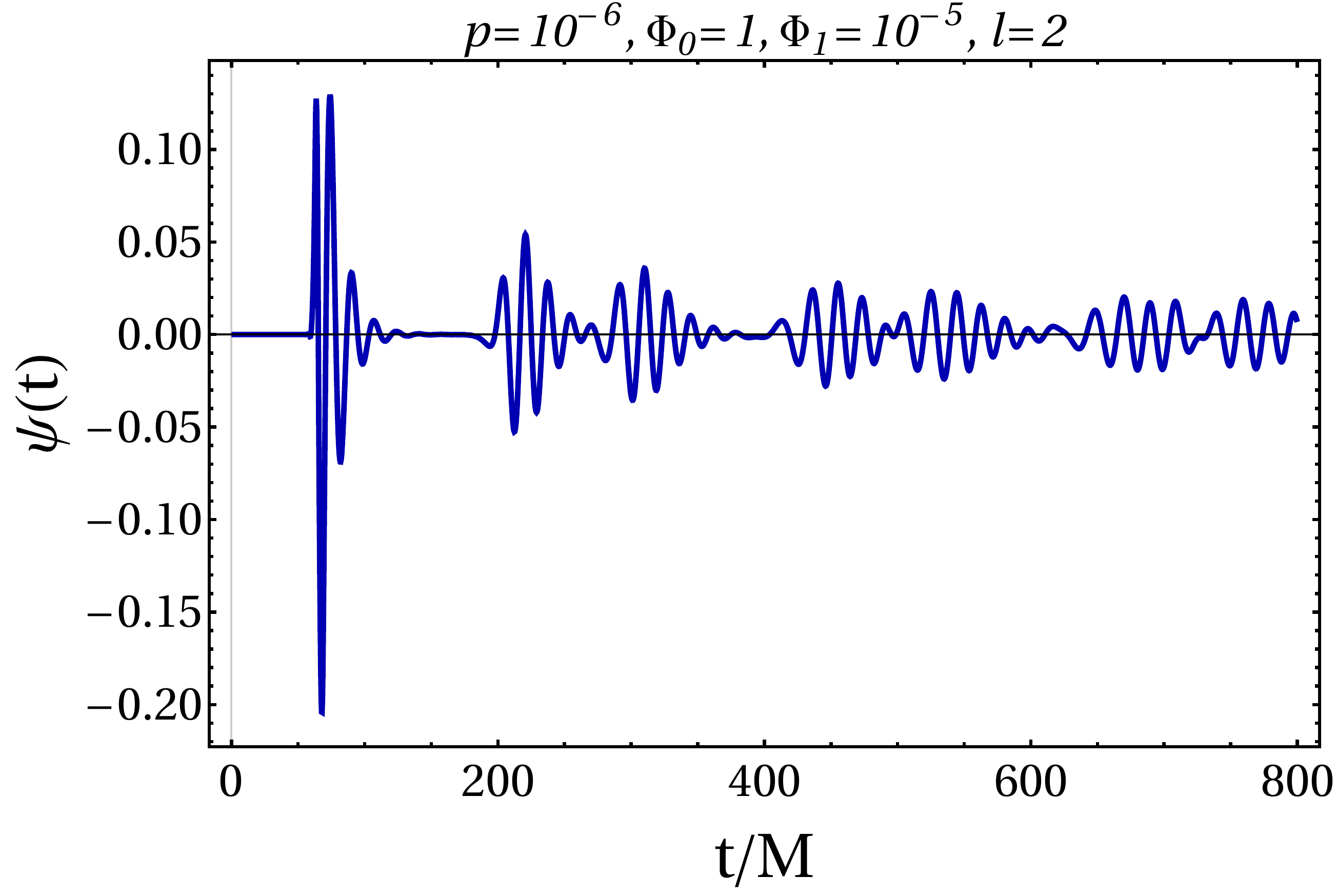}
	\endminipage\hfill
	\minipage{0.33\textwidth}
	\includegraphics[width=\linewidth]{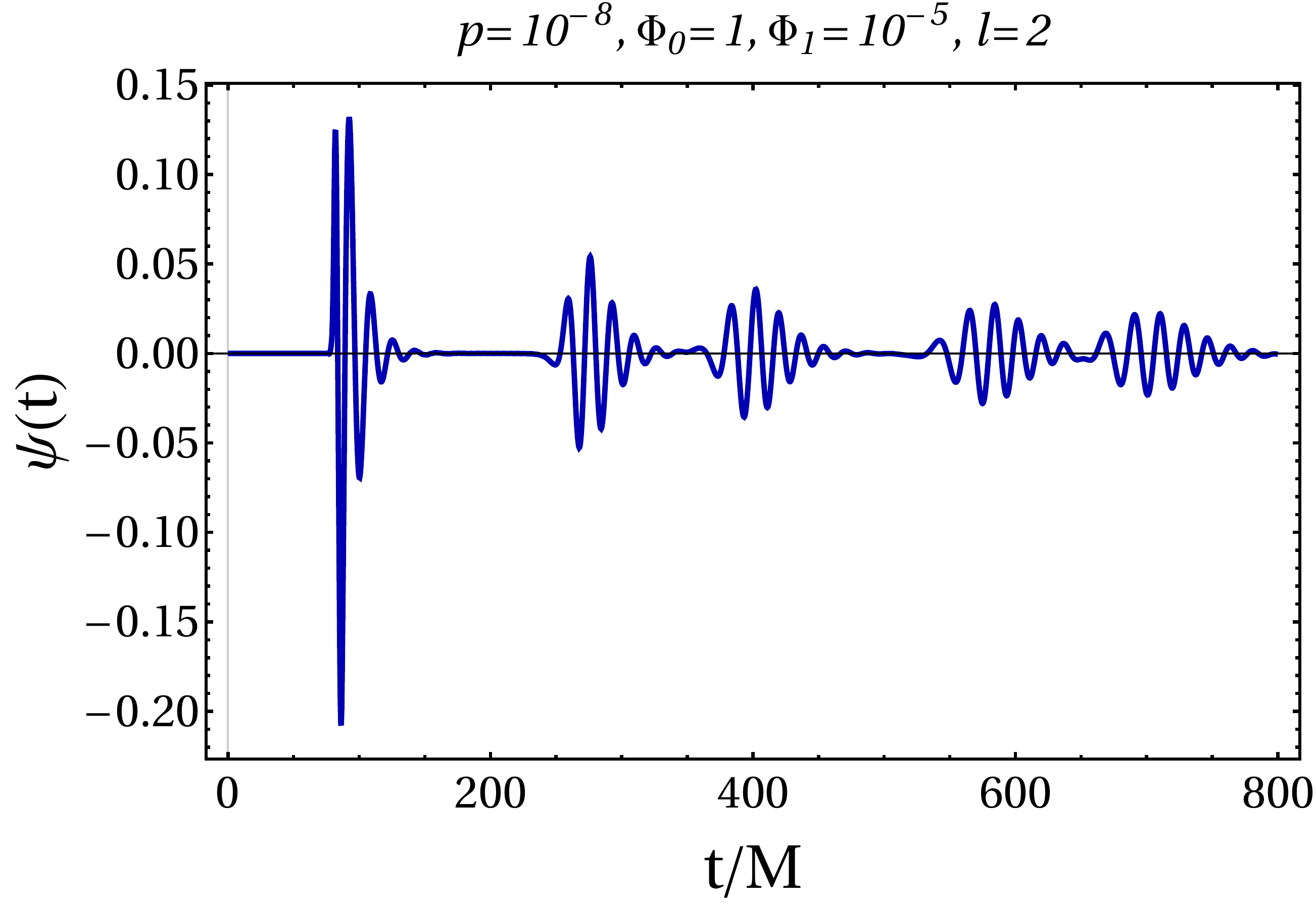}
	\endminipage\hfill
	\minipage{0.33\textwidth}%
	\includegraphics[width=\linewidth]{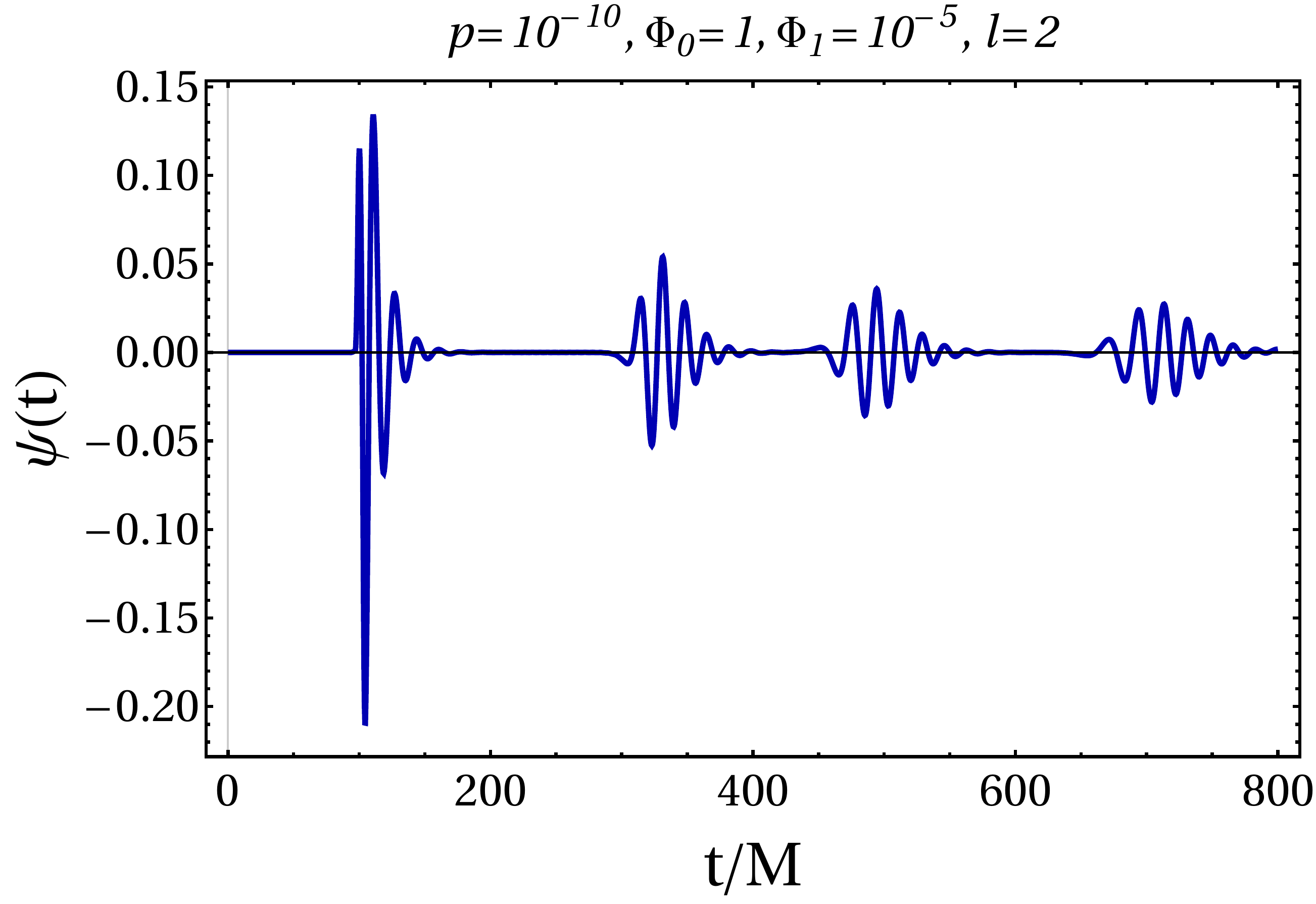}
	\endminipage
	\caption{The time-domain signals of the gravitational perturbation at $r=10M$, for $l=2$ and $p=10^{-6}$ (left panel), $p=10^{-8}$ (middle panel) and $p=10^{-10}$ (right panel), have been depicted. Here, we consider, $\Phi_0=1$ and $\Phi_1=10^{-5}$, appearing in the profile for the radion-like field $\Phi$ and use the Gaussian initial conditions given in \ref{i.c.} to obtain the ringdown signal. The behavior of the echo time, with the wormhole parameter $p$, continues to hold for gravitational perturbations as well.}\label{fig_ringdown_grav}
\end{figure}

In order to study the time-domain profile for the QNM $\psi_{lm}^{(s)}$, associated with the spin $s$ perturbation of the wormhole spacetime, we employ the inverse Fourier transformation in \ref{radial-master}. One example of such an inverse Fourier transform is the following, $-i\omega \psi_{lm~\textrm{(wormhole)}}^{(s)}\to\partial_t \hat{\psi}_{lm}^{(s)}$. Using such transformations, we obtain the following wavelike equation,
\begin{align}\label{wave_eqn}
\frac{\partial^2\hat{\psi}_{lm}^{(s)}}{\partial t^2}-\frac{\partial^{2}\hat{\psi}_{lm}^{(s)}}{\partial r_{*}^{2}}+V_{l~\textrm{(wormhole)}}^{(s)}(r)\hat{\psi}_{lm}^{(s)}=0~,
\end{align}  
with the outgoing boundary conditions being implemented as $\partial_{r_*}\hat{\psi}_{lm}^{(s)}= \mp \partial_{t}\hat{\psi}_{lm}^{(s)}$, as $r_*\to \pm \infty$. Here, we choose the initial conditions as \cite{Rahman:2021kwb}
\begin{equation}\label{i.c.}
\hat{\psi}_{lm}^{(s)}(0,r_*)=0,\qquad\partial_{t}\hat{\psi}_{lm}^{(s)}(0,r_*)=e^{-(r_*-7)^2}~,
\end{equation}
where $7$ is simply a convenient choice and it does not affect the results in any manner. With these boundary conditions, we solve \ref{wave_eqn} using the finite difference method. The results of such an analysis have been presented in \ref{fig_ringdown_Scalar} to \ref{fig_ringdown_grav}, respectively. In \ref{fig_ringdown_Scalar}, we show the time-domain profile of the scalar perturbation, with $l=2$ and for different values of the wormhole parameter $p$. In \ref{fig_ringdown_em} and \ref{fig_ringdown_grav}, we show the same for electromagnetic and gravitational perturbations. As evident from the plots, the primary signal is a black hole-like ringdown waveform, followed by a train of ever-modulated pulses known as the \textit{gravitational wave echoes} \cite{Cardoso:2017njb, Cardoso:2017cqb}. Moreover, the time separation between two successive pulses, the so-called the \textit{echo time} increases with the decrease of $p$. The phenomena can be understood in terms of the scattering of a Gaussian pulse off the double bump potential described in \ref{Model-V-l}. The potential forms a cavity of length $L$ in the tortoise coordinate. Now, consider a Gaussian pulse near the photon sphere of the wormhole (which coincides with the first bump in the geometric-optics limit). The fraction of pulse which scattered away to an asymptotic observer at infinity gives rise to the primary signal. The fraction of the pulse which leaks inside the cavity travels towards the second bump; part of which gets reflected back towards the photon sphere and gives rise to gravitational waves echo. The time taken by the pulse to complete this to-and-fro motion is given by $t_{\textrm{echo}}=2L\approx-16M \ln(p)$. This also explains why the echo time increases with the decrease of $p$. The reflection coefficient of the potential bump determines the modulation of the echo signal.    


We must mention another interesting point arising out of the time-domain signal. It is to be observed that irrespective of the nature of the perturbation, there are two consecutive echoes appearing in the time domain signal. It is not clear if this signal is due to the fact that the wormhole is taken to be symmetric around the throat, and hence there is a signal directly appearing from the potential on the other side of the throat as well, in addition to the one appearing due to reflection. In addition, note that as $p$ becomes smaller and smaller, i.e., the throat length becomes larger, this pairing gradually disappears. This suggests that possibly because for larger $p$, the throat length is small enough, such that the potentials on both sides of the wormhole have overlap and hence it can no longer be expressed as simply as \ref{Model-V-l}. Nonetheless, the origin of this unique signature needs to be understood better and whether this is unique for wormholes requires further study, which we postpone for a future work. 

\section{Possible generalization to rotating wormhole metric}\label{wormhole_rotation}

In the previous sections, we have observed that the imaginary parts of the QNM frequencies of the static and spherically symmetric braneworld wormhole, are very small. This is a signal of a potential instability, more so in the rotating context, which involves an ergoregion. The existence of negative energy states in the ergoregion can amplify these signals considerably, as these signals decay very slowly with time, owing to a very small imaginary part. To observe, if that is indeed the scenario, we consider generalizing the above static and spherically symmetric solution to the context of rotating spacetime as well. This can either be achieved by a Newman-Janis algorithm, as in \cite{newman1965note,Azreg-Ainou:2014pra}, or, by simple guesswork \cite{Bueno:2017hyj}. We will first present the more conventional Newman-Janis approach. 

For this purpose, we write the braneworld wormhole metric in terms of advanced Eddington-Finkelstein coordinates $\textrm(v,r,\theta, \phi)$, such that the metric in \ref{Sumanta-Form} reduces to
\begin{align}\label{metric-adv-SC}
ds^{2}=-\frac{1}{(p+1)^{2}}\left(p+\sqrt{1-\frac{2M}{r}}\right)^{2}dv^{2}+\frac{\left(p+\sqrt{1-\frac{2M}{r}}\right)2dvdr}{(p+1)\sqrt{1-\frac{2M}{r}}}
+r^{2}d\Omega^{2}~.
\end{align} 
The inverse metric components, resulting from the above line element can be written using the null-tetrad decomposition, $g^{\mu\nu}=-l^{\mu}n^{\nu}-n^{\mu}l^{\nu}+m^{\mu}\bar{m}^{\nu}+m^{\nu}\bar{m}^{\mu}$, where
\begin{align}
&l^{\mu}=-\frac{(p+1)\sqrt{1-\frac{2M}{r}}}{p+\sqrt{1-\frac{2M}{r}}}\delta^{\mu}_{r}~,\\
&n^{\mu}=\delta^{\mu}_{v}+\frac{1}{2}\sqrt{1-\frac{2M}{r}}\frac{\left(p+\sqrt{1-\frac{2M}{r}}\right)}{(p+1)}\delta^{\mu}_{r}~,\\
&m^{\mu}=\frac{1}{\sqrt{2}r}\left(\delta^{\mu}_{\theta}-\frac{i}{\sin\theta}\delta^{\mu}_{\phi}\right)~.
\end{align}
Subsequently, we allow $r$ to take complex values and replace the term $\frac{2}{r}$ with $\left[(1/r)+(1/\bar{r})\right]$. This is the conventional way to complexify the null-tetrads in the Newman-Janis algorithm. Then we perform the complex transformation $r\rightarrow r+ia\cos\theta$ and $v\rightarrow v+ia\cos\theta$ to get the desired rotating metric written in the null coordinates $\textrm(v,r,\theta, \phi)$, as,
\begin{equation}
g_{\mu\nu}^{\rm adv}=\frac{1}{K^{2}}
\begin{pmatrix}
-\left(1-\frac{2Mr}{\rho^{2}}\right)&K&0&-a\sin^{2}\theta\left(K-1+\frac{2Mr}{\rho^{2}}\right)\\
K&0&0&-aK\sin^{2}\theta \\
0&0&-(K\rho)^{2}&0\\
-a\sin^{2}\theta\left(K-1+\frac{2Mr}{\rho^{2}}\right)&-aK\sin^{2}\theta&0&\rho^{2}\Upsilon\sin^{2}\theta
\end{pmatrix}~,
\end{equation}
where, we have defined the quantities, $K$, $\rho^{2}$ and $\Upsilon$ for notational convenience, having the following expressions,
\begin{align}
K=\frac{(p+1)\sqrt{1-\frac{2Mr}{\rho^{2}}}}{p+\sqrt{1-\frac{2Mr}{\rho^{2}}}}~;\qquad \rho^{2}=r^{2}+a^{2}\cos\theta~;\qquad \Upsilon\equiv \left[\frac{a^{2}\sin^{2}\theta}{\rho^{2}}\left(1-\frac{2Mr}{\rho^{2}}\right)-\frac{2Ka^{2}\sin^{2}\theta}{\rho^{2}}-K^{2}\right]~.
\end{align}
Note that for $p=0$ we obtain the standard Kerr metric written in the advanced null coordinates. In order to bring the above metric in the standard Boyer-Lindquist coordinates, we must perform the following coordinate transformation,
\begin{align}
dv=dt+\sigma(r) dr~,
\label{t-equation}
\\
d\phi=d\chi +\gamma(r) dr~,
\end{align}
where, the integrability demands both $\sigma$ and $\gamma$ to be functions of the radial coordinate alone. We fix the form of functions $\sigma$ and $\gamma$, by requiring that the metric in the Boyer-Lindquist coordinate system $(t,r,\chi,\phi)$ must be symmetric under the transformation $\chi\rightarrow -\chi$ and $t\rightarrow -t$, which corresponds to $g_{rt}=0$ and $g_{r\chi}=0$. Using which, we obtain,
\begin{align}
\gamma=\frac{a}{r^{2}+a^{2}-2Mr}\equiv \frac{a}{\Delta}~,\\
\sigma=\frac{K\rho^{2}}{\Delta}+\frac{a^{2}\sin^{2}\theta}{\Delta}\label{btea-theta}~.
\end{align}
From \ref{btea-theta} we see that $\sigma$ explicitly depends on $\theta$, therefore \ref{t-equation} is not integrable to get a time-like coordinate $t$. On the other hand, for $p=0$, $K=1$ and $\sigma$ becomes a function of the radial coordinate alone. This is akin to the Kerr spacetime. Therefore, the conventional Newman-Janis algorithm fails to produce any rotating wormhole metric, starting from the static and spherically symmetric braneworld metric in the Boyer-Lindquist-like coordinate system.

This suggests looking for a rotating braneworld wormhole metric, from simple correspondence, as in \cite{Bueno:2017hyj}. Since for $p=0$ the metric in \ref{Sumanta-Form} reduces to that of Schwarzschild form, this inspires us to propose the following metric, 
\begin{align}
ds^{2}=-\frac{1}{(p+1)^{2}}\left(p+\sqrt{1-\frac{2Mr}{\rho^{2}}}\right)^{2}{}dt^{2}+\frac{\Sigma}{\rho^{2}}\sin^{2}\theta(d\phi-\omega dt)^{2}+\frac{\rho^{2}}{\Delta}dr^{2}+\rho^{2}d\theta^{2}~,
\end{align}
where, $\Sigma\equiv (r^{2}+a^{2})^{2}-a^{2}\Delta\sin^{2}\theta$ and $\omega=(2Mar/\Sigma)$. One can easily check that for $p=0$ this metric reduces to the Kerr metric and for $a=0$ it goes to \ref{Sumanta-Form}. As a future work, we may try to compute the energy-momentum tensor supporting this wormhole solution and we may investigate whether the above solution also allows for nonexotic matter fields. 

The above proposed rotating braneworld wormhole metric has another shortcoming, the metric is not separable, since both radial and $\theta$ coordinates appear under the square root. Thus the geodesic equation or even the Klein-Gordon equation cannot be expressed in a separable form. This may lead to the interesting phenomenon of resonance islands and would be something we wish to pursue in the future.

\section{Discussion and Concluding Remarks}\label{wormhole_discussion}

Wormholes are fascinating objects, connecting two distinct universes through a throat. The only problematic feature being, exotic matter fields are necessary in order for matter fields to travel through the throat between the two universes. Presence of extra dimensions cures this problem, since the higher spatial dimensions themselves mimic the role of exotic matter, while the "actual" matter fields on the four dimensional spacetime satisfy all the energy conditions. We have considered such a wormhole spacetime, known as the braneworld wormhole, where the length between the two branes acts as a real scalar field from the viewpoint of a four dimensional observer in the visible brane and the existence of a wormhole is intimately tied with the nontrivial behaviour of this interbrane separation. It will be interesting to also study the wormhole solutions in the single brane scenario \cite{PhysRevD.65.084040,Bronnikov:2002rn}, and compare them with the two-brane wormhole considered here. This will provide the necessary connection between perturbative and nonperturbative approaches in the context of the braneworld scenario.

Motivated by the unique perspectives offered by this static and spherically symmetric wormhole solution, namely it can mimic the ringdown signal of a black hole and at the same time can behave as an exotic compact object \emph{without} exotic matter, we have studied scalar, electromagnetic and axial gravitational perturbations on this wormhole geometry. In particular, we believe this is one of the first attempts(for earlier attempts see \cite{Roy:2021jjg,Blazquez-Salcedo:2018ipc}), where the gravitational perturbation of a wormhole spacetime has been discussed in detail. The resulting effective potential, governing the evolution of the gravitational perturbation, differs considerably from the potentials associated with the axial gravitational perturbation of black holes. This is because, the wormhole solution is a nonvacuum solution and one must take into account the perturbations in the matter sector as well. Using these effective potentials we provide the QNM frequencies for some of the lowest lying modes of scalar, electromagnetic, and gravitational perturbations, using both analytical and numerical prescriptions. It turns out that the real parts of the QNM frequencies derived using analytical methods match quite well with the numerical computation, but not so for the imaginary parts. Moreover, the imaginary parts of the QNM frequencies are very small, irrespective of the nature of the perturbations and these become even smaller, as the parameter $p$ becomes smaller. Thus we may argue that a larger value of $p\gtrsim 10^{-6}$ is necessary for the stability of the wormhole spacetime. This result, on the other hand, can possibly be verified in the future gravitational wave detectors, since a larger $p$ denotes a smaller length between the potentials present on the two sides of the throat and hence a smaller time delay. Thus the echoes will appear much closer to the primary ringdown signal, which has a much better chance of getting detected, if they exist, in the near future. In addition, if $p$ is too large $\gtrsim 10^{-1}$, then the echoes will almost coincide with the primary black hole-like ringdown spectrum, which can be ruled out given the gravitational waves measurements so far. Thus it seems that we have a conservative estimate of, $10^{-1}\gtrsim p\gtrsim 10^{-6}$, for the extra dimensional effect on the four dimensional brane, through $p$ (this is also supported by the recent black hole shadow measurement \cite{ShadowChak}). In turn, this provides a bound on the separation between the branes and hence the length of the extra dimensions as, $L=\ell\left(1+\ln \left[\ln q \right]\right)$, where, $0.2 \gtrsim q-1 \gtrsim 0.2\times 10^{-6}$ and $\ell$ is the characteristic length scale associated with the bulk cosmological constant. Thus our analysis can achieve two goals at the same time, first of all, it predicts what to expect from the gravitational wave signal, namely, very slowly decaying QNMs, leading to echoes and the associated echo time delay directly tells us about the nature of the extra dimensions. Moreover, we have observed that all the echoes, at least for the range of $p$ values considered above, comes in pairs, this is possibly another tell-tale signature of the existence of wormholes, if such echoes are indeed detected in the future gravitational wave detectors.

This work also has several future prospects to explore, e.g., such small values of the imaginary parts of the QNM frequencies can possibly signal instabilities, if some amplification mechanism is in place and this is precisely what happens for almost every rotating wormhole solution. To see if this is also the case here, it will be great to have a rotating braneworld wormhole solution and to study its stability under scalar, vector and tensor perturbations. Moreover, in the present case of static and spherically symmetric spacetime, we have not studied the case of polar gravitational perturbation, which can be done in a future project. Besides, the wormhole solution considered here corresponds to the two-brane scenario, which is perturbative in the bulk to brane curvature length scales. Thus, following \cite{Bronnikov:2019sbx}, it would be interesting to explore the gravitational wave signals in the nonperturbative regime, i.e., in the single brane scenario. In addition, one can work out possible effects if a wormhole inspirals a black hole. In particular, it will be interesting to ask about possible nonzero values of the tidal love number, multipole moments and also effects due to tidal heating on the wormhole spacetime. We hope to return to these issues in the near future.  
\section*{Acknowledgements}
Research of M.R. is funded by the National Post Doctoral Fellowship (N-PDF) from SERB, DST, Government of India (Reg. No.  PDF/2021/001234).
Research of S.C. is funded by the INSPIRE Faculty fellowship from DST, Government of India (Reg. No. DST/INSPIRE/04/2018/000893) and by the Start-Up Research Grant from SERB, DST, Government of India (Reg. No. SRG/2020/000409). S.B. and S.C. want to thank Ramit Dey for helpful discussions during the initial stages of this work. 
\appendix
\labelformat{section}{Appendix #1} 
\labelformat{subsection}{Appendix #1}
\section{Nature of the wormhole throat}\label{appB}

In this appendix, we will explicitly show that the discontinuity in the metric elements due to the presence of the square root term in the $g_{tt}$ component of the wormhole metric in \ref{S-kar-form} or, in \ref{Sumanta-Form} does not require an exotic surface stress tensor. For that purpose, we introduce a new coordinate $x$, such that $r-2M=x^{2}$. Under this transformation, the wormhole metric reduces to

\begin{align}
ds^{2}=-\frac{1}{(p+1)^{2}}\left(p+\frac{|x|}{\sqrt{2M+x^{2}}}\right)^{2}dt^{2}+4(2M+x^{2})dx^{2}+(2M+x^{2})^{2}(d\theta^{2}+\sin^{2}\theta d\phi^{2})~.
\end{align}
Let us now consider $x=\epsilon$ hypersurface, where $\epsilon$ is a constant, such that the limit $\epsilon \rightarrow 0$ yields the desired surface. First of all, note that the normalized normal to the $x=\epsilon$ hypersurface corresponds to, $n_{\alpha}=2\sqrt{2M+x^{2}}\,\nabla_{\alpha}x$, which is spacelike and hence the $x=\epsilon$ is the timelike hypersurface. As evident, $\partial_{x}g_{tt}$ involves derivative of $|x|$, yielding $\theta$-function, whose derivative would yield delta function at $x=0$ and hence we need surface stress tensor on that surface to account for such a behavior. Determining the induced metric is straightforward and can be obtained by first evaluating
\begin{align}
h_{\alpha \beta}=g_{\alpha \beta}-n_{\alpha}n_{\beta}=\textrm{diag.}\left(-\frac{1}{(p+1)^{2}}\left(p+\frac{|x|}{\sqrt{2M+x^{2}}}\right)^{2},0,(2M+x^{2})^{2},(2M+x^{2})^{2}\sin^{2}\theta \right)~,
\end{align}
whose projection on the $x=\epsilon$ hypersurface gives the induced metric. Given the induced metric, determination of the components of the extrinsic curvature are straightforward and can be derived using the relation, $K_{\alpha\beta}=h^{\mu}_{\alpha}\bigtriangledown_{\mu}n_{\beta}$. Since $n^{\alpha}K_{\alpha \beta}=0$, and $K_{\alpha \beta}$ is symmetric, the extrinsic curvature has only three nonzero components, yielding,
\begin{align}
K_{tt}&=-\Gamma^{x}_{tt}n_{x}~;
\qquad
\Gamma^{x}_{tt}=\frac{1}{4(2M+x^{2})(p+1)^{2}}\left(p+\frac{|x|}{\sqrt{2M+x^{2}}}\right)\frac{1}{\sqrt{2M+x^{2}}}\frac{x}{|x|}-\frac{x|x|}{(2M+x^{2})^{3/2}}~,
\\
K_{\theta\theta}&=x\sqrt{2M+x^{2}}=\frac{K_{\phi \phi}}{\sin^{2}\theta}~.
\end{align}
Here, we have used the result, $(d|x|/dx)=(x/|x|)=\theta(x)-\theta(-x)$, where the $\theta$-function is defined as, $\theta(x)=1$ for positive arguments and $\theta(x)=0$ for negative arguments. The trace of the extrinsic curvature can also be derived, either using the above components of the extrinsic curvature, or, directly from the normal to the $x=\epsilon$ hypersurface, 
\begin{align}
K=\nabla_{\alpha}n^{\alpha}=\frac{1}{\sqrt{-g}}\partial_{\alpha}\left(\sqrt{-g}n^{\alpha}\right)~.
\end{align}
The components of the normal vector have already been determined, while for the metric, its determinant becomes,
\begin{align}
\sqrt{-g}=\frac{2}{(p+1)}(2M+x^{2})^{3/2}\left(p+\frac{|x|}{\sqrt{2M+x^{2}}}\right)\sin\theta~,
\end{align}
and hence the extrinsic curvature turns out to be,
\begin{align}
K=\frac{x}{(2M+x^{2})^{3/2}}+\frac{1}{2\sqrt{2M+x^{2}}}\frac{1}{\left(p+\frac{|x|}{\sqrt{2M+x^{2}}}\right)}\left[\frac{1}{\sqrt{2M+x^{2}}}\frac{x}{|x|}-\frac{x|x|}{(2M+x^{2})^{3/2}}\right]~.
\end{align}
Given the components of the extrinsic curvature and its trace on the $x=\epsilon$ hypersurface, one can compute the following combinations, 
\begin{align}
\widetilde{K}^{t}_{t}&=K^{t}_{t}-h^{t}_{t}K=\frac{x}{(2M+x^{2})^{3/2}}~,
\\
\widetilde{K}^{\theta}_{\theta}&=K^{\theta}_{\theta}-h^{\theta}_{\theta}K
=-\frac{1}{2}\frac{1}{\sqrt{2M+x^{2}}}\frac{1}{\left(p+\frac{|x|}{\sqrt{2M+x^{2}}}\right)}\left[\frac{1}{\sqrt{2M+x^{2}}}\frac{x}{|x|}-\frac{x|x|}{(2M+x^{2})^{3/2}}\right]
=\widetilde{K}^{\phi}_{\phi}~.
\end{align}
From the above expressions it is easy to see that the jump of the above quantities across the $x=0$ hypersurface, defining the components of the surface stress-energy tensor, are given by,
\begin{align}
8\pi\widetilde{S}^{t}_{t}&\equiv \lim_{x\rightarrow 0^{-}}\widetilde{K}^{t}_{t}-\lim_{x\rightarrow 0^{+}}\widetilde{K}^{t}_{t}=0~,
\\
8\pi\widetilde{S}^{\theta}_{\theta}&\equiv \lim_{x\rightarrow 0^{-}}\tilde{K}^{\theta}_{\theta}-\lim_{x\rightarrow 0^{+}}\tilde{K}^{\theta}_{\theta}=\frac{1}{2Mp}
=8\pi\widetilde{S}^{\phi}_{\phi}
\end{align}
Therefore, it follows that the surface energy density $\sigma\equiv -\widetilde{S}^{t}_{t}=0$ and the isotropic surface pressure, $\mathbb{P}\equiv\widetilde{S}^{\theta}_{\theta}=(1/16\pi M p)$. Therefore $\sigma+\mathbb{P}>0$ at $r=2M$, and hence all the energy conditions are identically satisfied at the throat of the wormhole. 

\section{Connecting the tortoise coordinate of the wormhole spacetime with that of Schwarzschild black hole}\label{AppA}

In this appendix, we will provide the connection between the tortoise coordinate defined for the Schwarzschild black hole with that for the braneworld wormhole. For this purpose, we start with \ref{Sumanta-Form}, for which the tortoise coordinate $r_{*}$ will be defined through the following differential equation, $(dr/dr_{*})=\sqrt{-g_{tt}g^{rr}}$, whose integration yields,
\begin{align}
r_{*}=\int \frac{(p+1)dr}{\left(p+\sqrt{1-\frac{2M}{r}}\right)\left(1-\frac{2M}{r}\right)}+\textrm{constant}~.
\end{align}
Here, $p=(\alpha/\lambda)$ is the parameter inherited from extra spatial dimension. Imposing the boundary condition $r_{*}(r=2M)=0$ at the wormhole throat, located at $2M$, we obtain \cite{Aneesh:2018hlp}, 
\begin{align}
\frac{r_{*}}{1+p}=&\frac{M}{\lambda}\left[\frac{2(p-\beta)(2p-\beta)}{(p^{2}-1)[(p-\beta)^{2}-1]}+4\frac{\ln\frac{\beta}{p}}{(p^{2}-1)^{2}}+\frac{(p-2)\ln(1-p+\beta)}{(p-1)^{2}}-\frac{(2+p)\ln(1+p-\beta)}{(1+p)^{2}}\right]\\
=&\frac{r-2M}{\lambda(1-p^{2})}-\frac{pr\sqrt{1-\frac{2M}{r}}}{\lambda(1-p^{2})}+\frac{2M}{\lambda}\frac{\ln(\frac{r}{2M}-1)}{(1-p^{2})^{2}}
+\frac{4M}{\lambda}\left[\frac{\ln\left(1+\frac{p}{\sqrt{1-\frac{2M}{r}}}\right)-\ln p}{(1-p^{2})^{2}}\right]
\nonumber
\\
&\qquad \qquad +\frac{M}{\lambda}\frac{p^{3}-3p}{(1-p^{2})^{2}}\ln\left(\frac{1+\sqrt{1-\frac{2M}{r}}}{1-\sqrt{1-\frac{2M}{r}}}\right)~.
\end{align}
Since, the braneworld wormhole spacetime reduces to that of a Schwarzschild black hole in the limit $\alpha \rightarrow 0$ and $\lambda \rightarrow1$, we are interested in the parameter space $\alpha\gtrsim 0$ and $\lambda\lesssim 1$, i.e., for nearly Schwarzschild scenario. In this case, it is possible to approximate the above expression for the tortoise coordinate as,
\begin{align}
\frac{r_{*}}{1+p}&\approx\frac{r-2M}{1-(1-\lambda)}-\frac{pr}{1-(1-\lambda)}\sqrt{1-\frac{2M}{r}}+\frac{2M}{1-(1-\lambda)}\ln\left(\frac{r}{2M}-1\right)
\nonumber
\\
&\qquad +\frac{4M}{1-(1-\lambda)}\left[\ln\left(1+\frac{p}{\sqrt{1-\frac{2M}{r}}}\right)-\ln p\right]
+\frac{M}{1-(1-\lambda)}(-3p)\ln\left(\frac{1+\sqrt{1-\frac{2M}{r}}}{1-\sqrt{1-\frac{2M}{r}}}\right)~.
\end{align}
Note that we have expressed the term involving $\lambda$ in the denominator, as $\lambda=1-(1-\lambda)$, so that we can use the approximated result, $[1-(1-\lambda)]^{-1}\approx 1+(1-\lambda)$, since $(1-\lambda)$ is a small quantity. Using this approximation and neglecting all the second order infinitesimals like, $p(1-\lambda)$, $(1-\lambda)^{2}$, and $p^{2}$, we finally obtain,
\begin{align}\label{Tortoise-5}
r_{*}=&\left(1+p\right)\left[r+2M\ln\left(\frac{r}{2M}-1\right)\right]+2M\left(1+p\right)\left[-1-2\ln p\right]+(1-\lambda)\left[(r-2M)+2M\ln\left(\frac{r}{2M}-1\right)\right]
\nonumber
\\
&\qquad +p\left[-r\sqrt{1-\frac{2M}{r}}+\frac{4M}{\sqrt{1-\frac{2M}{r}}}-3M\ln\left(\frac{1+\sqrt{1-\frac{2M}{r}}}{1-\sqrt{1-\frac{2M}{r}}}\right)\right]-4M(1-\lambda)\ln p~.
\end{align}
Introducing the tortoise coordinate for the Schwarzschild black hole spacetime as, $r_{*}^{\rm BH}\equiv r+2M\ln\left[(r/2M)-1\right]$, we will arrive at \ref{tortoise-3}, used in the main text. 

\bibliography{Reference_1}

\bibliographystyle{./utphys1}
\end{document}